\documentclass[12pt]{article}
\usepackage[body={17cm, 23cm, centered}]{geometry}

\usepackage[utf8]{inputenc}
\usepackage[T1]{fontenc}

\usepackage[unicode]{hyperref}
\hypersetup{colorlinks=true,linktocpage,breaklinks,
            urlcolor=blue,
            linkcolor=red,
            citecolor=blue
            }

\usepackage{amsmath,amssymb,amsthm,amscd,cite,comment,amsfonts,indentfirst,color,setspace,bbold}
\usepackage{mathtext,marvosym,textcomp}
\usepackage{arydshln}
\usepackage[makeroom]{cancel}

\usepackage{tocloft}

\usepackage{subcaption}

\usepackage{epsfig,verbatim,cite}
\usepackage{braket}

\usepackage{mathtools}
\usepackage{slashed}
\usepackage[usenames,dvipsnames,svgnames,table]{xcolor}
\usepackage{tikz}
\usepackage[ normalem]{ulem}

\usepackage{float}

\usepackage{multirow}

 \usepackage{relsize}

 \DeclareMathOperator*{\Sum}{\mathlarger{\mathlarger{\sum}}}
 \DeclareMathOperator*{\SSum}{\mathlarger{\mathlarger{\mathlarger{\sum}}}}

\numberwithin{equation}{section}

\def\a{\alpha} \def\b{\beta} \def\g{\gamma} \def\d{\delta} \def\e{\epsilon}
 \def\z{\zeta} \def\h{\eta} \def\q{\theta}
  \def\k{\kappa} \def\m{\mu}
\def\n{\nu} \def\x{\xi} \def\p{\pi}  \def\r{\rho}
 \def\s{\sigma}   \def\f{\phi}
  \def\y{\psi}

\def\G{\Gamma} 
\def\D{\Delta}

 \def\S{\Sigma}  
\def\F{\Phi}

\def\fr{\frac} \def\dfr{\dfrac} \def\dt{\partial}

\def\ph{\phantom}
\def\mc{\mathcal}

\def\SS{\mathbb{S}}
\def\HH{\mathbb{H}}
\def\ZZ{\mathbb{Z}}

\def\rmU{\mathrm{U}}
\def\rmSU{\mathrm{SU}}

\def\rmSO{\mathrm{SO}}
\def\rmE{\mathrm{E}}

\newcommand\bqa {\begin{eqnarray}}
\newcommand\eqa {\end{eqnarray}}

\newcommand{\bear}{\begin{array}}
\newcommand{\enar}{\end{array}}

\newcommand{\cM}{\mathcal{M}}

\newcommand{\be}{\begin{equation}}
\newcommand{\ee}{\end{equation}}
\newcommand{\bea}{\begin{eqnarray}}
\newcommand{\eea}{\end{eqnarray}}

\usepackage{tikz}
\usetikzlibrary{arrows}
\usetikzlibrary{shapes.misc}
\usepackage{tikz-cd}

\usetikzlibrary{arrows,shapes,positioning, fit}
\tikzstyle{every picture}+=[remember picture]
\tikzstyle{na} = [baseline=-.5ex]
\tikzstyle{format} = [rounded rectangle,
                      thick,
                      minimum size=1cm,
                      draw=red!00!black!80,
                      top color=white,
                      bottom color=red!00!black!00,
                      on grid]
\tikzstyle{format1} = [rectangle,
                      thick,
                      minimum size=1cm,
                      draw=red!00!black!80,
                      top color=white,
                      bottom color=red!00!black!00,
                      on grid]
\tikzstyle{format0} = [rounded rectangle,
					  thick,
				      minimum size=1.2cm,
					  draw=blue!00!black!80,
					  top color=white,
                      bottom color=blue!00!black!00,
                      text centered]
\tikzstyle{formatd} = [rounded rectangle,
					  thick,
					  dashed,
				      minimum size=1cm,
					  draw=blue!50!black!80,
					  top color=white,
                      bottom color=blue!00!black!00,
                      text centered]
\tikzstyle{format1d} = [rounded rectangle,
                      thick,
                      dashed,
                      minimum size=2cm,
					  draw=blue!50!black!80,
					  top color=white,
                      bottom color=blue!00!black!00,                  
                      text centered]
                      
\tikzset{cross/.style={cross out, draw=black, minimum size=2*(#1-\pgflinewidth), inner sep=0pt, outer sep=0pt},
	cross/.default={5pt}}
\usepackage[ normalem]{ulem}
\usepackage{tocloft}

\begin{document}
\renewcommand{\contentsname}{}
\renewcommand{\refname}{\begin{center}References\end{center}}
\renewcommand{\abstractname}{\begin{center}\footnotesize{\bf Abstract}\end{center}} 

\begin{titlepage}
\ph{preprint}

\vfill

\begin{center}
   \baselineskip=16pt
   {\large \bf Holographic RG flows and boundary conditions \\ in a 3D gauged supergravity
   }
   \vskip 2cm
    Ksenia Arkhipova$^{a}$\footnote{\tt arkhipova@theor.jinr.ru},
   Lev Astrakhantsev$^{b}$\footnote{\tt lev.astrakhantsev@phystech.edu},
    Nihat Sadik Deger$^c$\footnote{\tt sadik.deger@boun.edu.tr}, \\
    Anastasia A. Golubtsova$^a$\footnote{\tt golubtsova@theor.jinr.ru },
    Kirill Gubarev$^b$\footnote{\tt kirill.gubarev@phystech.edu },
    Edvard T. Musaev$^{b,d}$\footnote{\tt musaev.et@phystech.edu}	
       \vskip .6cm
             \begin{small}
                          {\it
                          $^a$Bogoliubov Laboratory of Theoretical Physics, JINR, Joliot-Curie str. 6, Dubna, 141980 Russia\\ 
                          $^b$Moscow Institute of Physics and Technology, Dolgoprudny, 141702, Russia \\
                          $^c$Department of Mathematics, Bogazici University, Bebek, 34342, Istanbul, Turkey,\\
                          $^d$Institute for Theoretical and Mathematical Physics, MSU, Moscow 119991, Russia                 
                          } \\ 
\end{small}
\end{center}

\vfill 
\begin{center} 
\textbf{Abstract}
\end{center} 
\begin{quote}
\small
In this work we focus on the study of RG flows of conformal field theories that are holographically dual to Poincar\'e domain wall solutions in $D=3$, $\mc{N}=(2,0)$ gauged supergravity coupled to a sigma model with target space 
$\rmSU(1, 1)/\rmU(1) = \HH^2$. This theory is truncated to a subsector where the vector field and phase of the scalar field vanish and we consider different boundary conditions for the remaining real scalar field. The RG flows, which are mostly non-superysymmetric, are analyzed by treating the supergravity field equations as a dynamical system for the scalar field and its derivative with respect to the scale factor. Phase diagrams are constructed for different values of the parameter $a^2$, which is related to the curvature of the scalar manifold. The behavior of solutions near the boundary is used to determine their type based on the expansion of the corresponding fake superpotential. By incorporating information on the boundary conditions, the obtained RG flows are interpreted using the holographic dictionary. Numerical solutions and plots of the fake superpotential are also provided.
\end{quote}

\vfill
\setcounter{footnote}{0}
\end{titlepage}

\tableofcontents

\setcounter{page}{2}

\section{Introduction}
\label{sec:intro}

In contrast to classical field theories defined by a Lagrangian or field equations, definition of a quantum field theory necessarily includes an energy scale. After a renormalization procedure, the effective Lagrangian of a quantum field theory including counterterms depends on this scale in a way prescribed by the renormalization group equations. In the standard approach, taking into account loop corrections, dependence on an arbitrary dimensionful  parameter $\mu$ is introduced, which must not enter in the final result for an amplitude. For that a reference energy scale $E_*$ (or a certain position in the momentum space of external legs) must be chosen where coupling constants are compared to experiment and define the rest of the RG flow. It is often convenient to choose this point to be the critical UV point of the theory where symmetry gets maximally enhanced or the theory becomes conformal. In the latter case the RG flow from the UV critical point is triggered by an IR relevant operator or a non-vanishing VEV of a field of the theory thus breaking the symmetry. In principle, the symmetry might be restored at an IR critical point.

In the perturbative approach to quantum field theory, renormalization group flow of a coupling constant $g$ can be derived by calculating the scattering matrix, or more precisely vertex functions, up to a given loop order (see e.g. \cite{Gell-Mann:1954yli,Bogolyubov:1956gh,Bogolyubov:1959bfo}). Regularization of UV divergences coming from loops and further renormalizations result in the well-known Callan-Symanzik equation for the beta function \cite{Callan:1972uj,Symanzik:1970rt}
\begin{equation}
    \m \fr{\dt g}{\dt \m} = \b_g(g,\m).
\end{equation}
In this process usually certain amount of information about high energy behavior of the QFT couplings is lost, which in contrast is preserved in the related but conceptually different methods of the Wilsonian \cite{Wilson:1974mb} renormalization group and Polchinski equation \cite{Polchinski:1983gv}. 

A fundamentally different approach to (non-perturbative) construction of renormalization group flows of a QFT is based on the (weak) holographic duality between a non-gravitational quantum field theory and classical supergravity equations. The AdS/CFT duality idea \cite{Maldacena:1997re} was inspired by the observation that BPS states of the Type IIB string theory on  AdS${}_5\times \SS^5$ at weak coupling matches with that of the $\mc{N}=4$, $D=4$ SYM at strong coupling and it was made more precise later in \cite{Gubser:1998bc,Witten:1998qj}. The proposed algorithm to obtain Green's functions of $\mc{N}=4$, $D=4$ SYM is based on the idea that the effective action of the gauge theory is equal to the classical action of the gravitational theory living on the conformal boundary of AdS. The precise correspondence has been conjectured to follow a simple equation:
\begin{equation}
    \left\langle e^{\int_{\dt M} \phi_0 \mc{O}}\right\rangle = \mc{Z}[\f_0],
\end{equation}
where the partition function of the gravitational theory on the RHS becomes a function of the classical action in the saddle point approximation. For the $\mc{N}=4$ SYM theory on the LHS this corresponds to taking t'Hooft limit, thus reproducing the observation of \cite{Maldacena:1997re}. Based on these results in \cite{Akhmedov:1998vf} has been shown that equations of motion of Type IIB supergravity on AdS${}_5 \times \SS^5$ are equivalent to renormalization group flow equations of the dual gauge theory and a general relation between second order field equations and first order RG flow equations has been proposed (see also the review \cite{Akhmedov:2001un}). 

In the holographic approach to renormalization group flow, UV divergences on the QFT side correspond to near-boundary divergences on the gravitational side. Since in quantum field theories UV divergences do not depend on the IR physics, on the gravitational side the holographic renormalization should involve only the near-boundary analysis. The formalism first introduced in \cite{Henningson:1998gx,Akhmedov:1998vf} and promoted to a systematic procedure in \cite{deHaro:2000vlm} automatically incorporates Ward identities and their anomalies in the form of kinematical constraints. Dynamical information on the QFT, i.e. the correlation functions, is then encoded in geometry on the gravitational side (for a review see \cite{Skenderis:2002wp,Papadimitriou:2016yit}). A direct correspondence between supergravity equations of motion and renormalization group equations has been described later in \cite{deBoer:1999tgo} and \cite{Martelli:2002sp} in terms of descent equations following from the Hamiltonian-Jacobi method where the radial direction plays the role of the evolution parameter (for a review see \cite{deBoer:2000cz,Papadimitriou:2003is,Northe:2022tqq}). At the same time on the QFT side Callan-Symanzik-Polchinski equations have been written in the Hamiltonian form in \cite{Akhmedov:2002gq} (see also later results \cite{Akhmedov:2010mz,Akhmedov:2010sw} for relation to integrability).  Dealing with canonical variables for the bulk theory instead of the second order field equations allowed to develop a covariant approach using eigenfunctions of the dilatation operator in \cite{Papadimitriou:2004ap}. 

As in the standard renormalization procedure in a QFT,  counterterms to the supergravity Hamiltonian are of crucial importance to remove divergences on the boundary. The Gibbons-Hawking-York term alone is not enough for this and one has to search for counterterms that i) remove all divergences and ii) are formulated in covariant terms, say covariant w.r.t. the dilatation operator exploiting the notion of scale. A procedure of computing the finite (renormalized) stress-energy tensor based on adding a covariant functional of the intrinsic boundary geometry has been proposed in \cite{Balasubramanian:1999re,Kraus:1999di}. The procedure is intrinsic to the geometry, unambiguous once the counterterm is specified and is completely analogous to removal of divergences in quantum field theory. Further in \cite{deHaro:2000vlm} a systematic procedure of renormalization for the AdS/CFT prescription has been developed, that can be understood as the holographic reconstruction of the bulk space-time from the boundary CFT data. The idea is to turn to the Fefferman-Graham coordinate system 
\begin{equation}
    ds^2 = \fr{l^2}{r^2}\left(dr^2 + g_{ij} (x, r)dx^i dx^j \right),
\end{equation}
and expand all variables in powers of $r$. Counterterms are then determined by terms with particular powers (see Appendix \ref{app:renorm} for the simplest case of a scalar field on the AdS). One- and two-point correlation points have been computed using this holographic renormalization procedure in \cite{Bianchi:2001de} for Poincar\'e domain wall solutions, where each $r=$const. slice has Poincar\'e symmetry. A systematic procedure for deriving renormalized on-shell action for Poincar\'e domain walls with AdS boundary has been developed in \cite{Bianchi:2001kw} for Dirichlet boundary conditions and applied to the maximal D=5 gauged supergravity. 

In this paper we apply the covariant holographic renormalization procedure to the on-shell action of the $\mc{N}=(2,0)$, $D=3$ gauged supergravity of \cite{Deger:1999st,Deger:2002hv} coupled to a single scalar multiplet where the scalar manifold is  $$\fr{\rmSU(1, 1)}{\rmU(1)} = \HH^2.$$ We focus at Poincar\'e domain wall solutions with asymptotically AdS boundary, holographically describing renormalization group flows from a conformal critical point. Following \cite{Papadimitriou:2007sj} and \cite{Papadimitriou:2004rz} we consider field configurations both with Dirichlet and mixed boundary conditions (that include Neumann b.c. as a particular case). As it has first been observed in \cite{Klebanov:1999tb}, a given field configuration may correspond to two different CFT's depending on the boundary conditions chosen, that results in different behavior of correlators on the boundary. For completeness we briefly review this in Appendix \ref{app:renorm}. Although an exact domain wall solution to the field equations of the theory has been found in \cite{Deger:1999st}, in general one cannot provide a family of exact solutions. This is a half-supersymmetric solution and the RG flows that it describes were studied in \cite{Deger:2002hv} as well as a particular non-supersymmetric flow between 2 different AdS vacua. Here we extend that work to a more general family of solutions by relaxing the supersymmetry requirement and allowing different boundary conditions.
To do that we follow the near-boundary analysis and the covariant renormalization technique which we briefly review in Appendix \ref{app:covren} and in Section \ref{sec:renfunc} below. In the absence of exact solutions we analyze renormalization group flows of the dual CFT using the dynamical system interpretation suggested in \cite{Gukov:2016tnp} and further developed in \cite{Arefeva:2018jyu,Arefeva:2018hyo,Arefeva:2019qen,Arefeva:2020aan,Arav:2020asu}, where RG flows appear as phase trajectories in the variables corresponding to VEV of the scalar field and its beta-function. The analysis presented here completes that of \cite{Golubtsova:2022hfk} performed for the same model by incorporating all possible boundary conditions and all possible types of solutions, covering both RG flows triggered by a non-zero VEV and by adding a relevant operator. We also find exotic renormalization group flows, that seem to be pretty standard from the gravitational side but have been little understood in QFT terms (see also \cite{Kiritsis:2016kog,Ghosh:2017big,Gursoy:2018umf}). 

The paper is organized as follows: In Section \ref{sec:model} we introduce the model of $\mc{N}=(2,0)$, $D=3$ gauged supergravity of interest, derive its first order equations based on the notion of fake superpotential and discuss derivation of correlation functions for various boundary conditions. In Section \ref{sec:sols} we discuss near-boundary behavior of the fields of the model in the Poincar\'e domain wall ansatz, derive dynamical system equations describing holographic RG flows and provide their phase pictures. Among these we detect exotic flows corresponding to bouncing solutions and comment on their interpretation and meaning. We conclude in Section \ref{sec:disc} where we discuss our main results and sketch further directions to explore. In Appendix \ref{appendixA} we give a short review of the boundary value problem for scalar fields on AdS  and in Appendix \ref{app:autonom} we show the derivation of the dynamical system equations \eqref{eqZX} that we use.

\section{The model: a 3D gauged supergravity}
\label{sec:model}

At the gravity side the model will be taken to be the $\mc{N}=(2,0)$  AdS$_3$ supergravity coupled to $n$-copies of $\mc{N}=(2,0)$ scalar multiplet constructed in \cite{Deger:1999st}. The field content of this 3-dimensional supergravity is given by a vielbein $e_\m{}^a$, a doublet of gravitini $\y_\m$ and a gauge field $A_\m$. The scalar multiplet consists of a complex scalar field $\F$ and by a doublet of spinorial fields $\lambda$ where R-symmetry indices are omitted. In \cite{Deger:1999st} the case where scalars parametrize a coset space $G/H$ was considered where $G$ can be compact or non-compact and $H$ is its maximal compact subgroup. In this paper we will only consider the non-compact case for $G$ with a single complex scalar field $\F$
(that is, $n=1$) where the coset space is   
\begin{equation}
    \begin{aligned}
        \fr{\rmSU(1,1)}{\rmU(1)} = \HH^2.
    \end{aligned}
\end{equation}
In this case the bosonic part of the supersymmetric Lagrangian is \cite{Deger:1999st}
\begin{equation} \label{lag1}
    e^{-1}\mc{L} = \fr14 R - \fr{e^{-1}}{16 m \, a^4} \e^{\m\n\r}A_\m \dt_\n A_\r - \fr{|D_\m \F|^2}{a^2(1- |\F|^2)} -   V(\F), 
\end{equation}
where $e = \det e_\m{}^a$ and $D_\mu\F= (\partial_\mu+iA_\mu)\F$.
 Here $-4m^2$ is the AdS$_3$ cosmological constant
and the
constant $a$  is related to the curvature of the hyperbolic scalar manifold and is non-zero\footnote{The $a=0$ case corresponds to the flat sigma model. To take this limit, first fields $\phi$ and $A_\mu$ have to be rescaled with appropriate powers of $a$ and after the limit the potential in the action reduces to a cosmological constant  \cite{Deger:2000as,Izquierdo:1994jz}.}. The potential $V(\F)$ is given by
\begin{equation}
    V(\F) =  2 {m^2} C^2 \left(2 a^2 |S|^2 -  C^2 \right) \, ,
\label{pot}
\end{equation}
where
\begin{equation}
    C = \fr{1 + |\F|^2}{1 -  |\F|^2}, \quad     S = \fr{2 \F}{1 - |\F|^2}.
\end{equation}

To bring the kinetic term of the scalar field to the standard form one splits the complex scalar field $\F$ into its modulus $|\F| \equiv \phi$ and phase $\q$. For the former we additionally define
\begin{equation}
        C \equiv \cosh \f \,\,, \,\, |S| \equiv \sinh \f .
\end{equation}
After this, the Lagrangian of the theory becomes
\begin{equation}
\label{fullmodel}
    \begin{aligned}
    e^{-1}\mc{L} &= \fr14 R - \fr{e^{-1}}{a^4} \e^{\m\n\r}A_\m \dt_\n A_\r - \fr1{4a^2} \dt_\m \f \dt^\m \f\\
    &- \fr1{4a^2} |S|^2 (\dt_\m \q + A_\m)(\dt^\m \q + A^\m) -  V(\f).
    \end{aligned}
\end{equation}
Note that the potential depends only on the field $\f$.
From the field equations of $\q$ and the vector field $A_\m$, it is easy to see that setting
$A_\m= \q = 0$ is a consistent truncation of the theory \cite{Deger:1999st}, which finally brings us to the action of the form
\begin{equation}
    \label{eq:sugra}
        S = \frac{1}{4} \int d^{3}x\sqrt{-g}\left(R -\frac{1}{a^2}(\partial\phi)^2-{4}V(\phi)\right).
\end{equation}

We now proceed with the remaining field equations of the theory
after the truncation. The Einstein equations of \eqref{eq:sugra} read
\begin{equation}
R_{\mu\nu} -\frac{1}{2}g_{\mu\nu} R= T_{\mu\nu},
\end{equation}
where the stress energy tensor takes the usual form
\begin{equation}
 T_{\mu\nu}=\frac{1}{a^2}\left(\partial_{\mu}\phi\partial_{\nu}\phi-\frac{1}{2}g_{\mu\nu}\partial_{\sigma}\phi\partial^{\sigma}\phi\right)-2g_{\mu\nu}V(\f) .
\end{equation}
For the scalar field we have
\begin{equation}
    \frac{1}{\sqrt{-g}}\dt_\m \left(\sqrt{-g} g^{\m\n} \dt_\n \f \right) - 2 a^2V'(\f)=0,
    \label{scalar}
\end{equation}
which is the standard Klein-Gordon equation where prime indicates derivative w.r.t. the field $\phi$, that is $V'\equiv \partial_{\phi} V $. Given the intended application to holographic renormalization group flows we are interested in Poincar\'e domain wall solutions, that is the metric will be restricted to the following ansatz
\begin{equation}\label{metricT}
ds^2=e^{2A(r)}\left(-dt^2+dx^2\right)+dr^2.
\end{equation}
Here $r$ is a radial coordinate transverse to the domain wall and is defined on the interval  $(0, +\infty)$. The scalar field will also have only dependence on the radial coordinate $\phi= \phi(r)$.
Then, the Einstein equations together with the scalar field equations may be written as follows \cite{Deger:2002hv}
\begin{eqnarray}
2\ddot{A}+2\dot{A}^2+{4}V+\frac{\dot{\phi}^2}{a^2}&=&0,\label{eom1}\\
2\dot{A}^2+{4}V-\frac{\dot{\phi}^2}{a^2}&=&0,\label{eom3}\\
\ddot{\phi}+ 2\dot{A}\dot{\phi}- 2a^2 V' &=& 0,
\label{eqd}
\end{eqnarray}
where dot denotes derivative w.r.t. the radial coordinate, e.g. $\dot \f \equiv \dt_r \f$. Only two equations are independent. From the equations (\ref{eom1})-(\ref{eom3}) one also finds the relation
\begin{equation}\label{eomextra}
\ddot{A}+\frac{\dot{\phi}^2}{a^2}=0,
\end{equation}
that will be useful in what follows. It is easy to check that when the potential \eqref{pot} is written as (which is always possible due to Hamilton-Jacobi theory)
\begin{equation}
\label{eq:V2W}
    V = \fr{a^2}{4}W'^2 - \fr12 W^2 ,
\end{equation}
then,
the following first order equations \cite{Deger:2002hv}
\begin{equation} \label{first}
    \begin{aligned}
        \dot A &= - W, \\
        \dot \f &= a^2 W', 
    \end{aligned}
\end{equation}
solve the above second order field equations. It is important to notice that the differential equation \eqref{eq:V2W} has in general more than one solution. Only one in this solution set is the actual (true) superpotential and the corresponding field configurations will be supersymmetric by construction. Following \cite{Papadimitriou:2007sj} we will refer to all the others as fake superpotentials. These in general correspond to non-supersymmetric solutions. From the supersymmetry variations of the model \cite{Deger:1999st}, the superpotential of this model was found to be \cite{Deger:2002hv}:
\begin{equation}
    W_{\rm susy} = - 2 {m} \cosh^2 \phi.
\end{equation}

If a critical point of $V(\f)$ is also a critical point of $W(\f)$, then it is a gravitationally stable  critical point. Depending on whether the scalar field behaves near the boundary as $e^{-2m\, \D_+ r}$ or $e^{-2m\, \D_- r}$, the superpotential has two possible Taylor series expansions around a critical point $\phi_*$ of the potential, located say at $\phi=0$:
\begin{equation} \label{wpm1}
    W_\pm (\f) = -2{m}\left(1 + \frac{1}{2a^2} \D_\pm \f^2 + ...\right).
\end{equation}
These are simply two solutions to the equation \eqref{eq:V2W} up to the order $\f^2$ with the condition $W(\f_*) < 0$. Note that the actual behavior of the superpotential is determined by the explicit form of the solution, to which we will refer as $W_+$ and $W_-$ type solutions.

\subsection{Renormalized generating functionals}
\label{sec:renfunc}

Fake superpotential is  an important ingredient in the holographic renormalization procedure based on the minisuperspace approach, where all moduli of the background domain wall metric \eqref{metricT} are frozen except the factor ruled by $A(r)$. In fact fake superpotential gives precisely the on-shell action for the domain wall ansatz and hence the effective action (up to renormalization) of the dual quantum field theory. The domain wall solutions that we consider have asymptotically locally AdS behavior (AlAdS) at large $r$ and hence, possess a conformal boundary as $r \to \infty$. In this region the standard analysis of a scalar field on the AdS space-time can be applied to see that the on-shell action diverges. Then once can introduce a regulating surface $\S_r$ at some finite $r$, renormalize physical quantities (correlators) and then send $r \to \infty$. From the point of view of the dual field theory this finite $r$ corresponds to a regulating UV cut-off. There is however a subtlety that not all boundary conditions defined on $\S_r$ will make sense when the regulator is removed. Indeed, one is able to introduce an action with a radial cutoff that is compatible with boundary conditions at $\S_r$ while to define a standard dual CFT boundary conditions must be imposed at the conformal boundary. However, boundary conditions at a finite radial cutoff are not locally related to those at the conformal boundary and hence do not define a dual CFT in the standard prescription. In our analysis we will be following the holographic renormalization approach of \cite{Papadimitriou:2004ap} and \cite{Papadimitriou:2005ii} that takes into account this subtlety by decomposing fields into irreps of the boundary conformal group. In particular in \cite{Papadimitriou:2005ii} it has been shown (for Dirichlet boundary conditions) that the covariant local counterterms are necessary to make the variational problem well-posed. As a consequence the resulting action is finite. We illustrate this approach for a scalar field on a fixed AdS background in Appendix \ref{app:scalarads} and \ref{app:renorm}.

As it has been discussed in Section \ref{sec:intro} dual field theory interpretation of a given classical supergravity solution depends on the choice of boundary conditions near the AlAdS boundary. More strictly, to consistently define a field theory on a space-time with a boundary one must fix not only the bulk action responsible for field equations, but also the boundary action that correctly takes into account boundary conditions for field variations. For pure gravity with Dirichlet boundary conditions for metric variations, i.e. $\d g_{\m\n}|_{\S_r}  = 0$, the boundary action is given by the well-known Gibbons-Hawking-York term given by the trace of the extrinsic curvature $K$, see Appendix \ref{app:covren}. Hence, including the boundary term for the scalar field $\f$ we have the following action for our model
\begin{equation}
    \label{eq:action0}
    \begin{aligned}
    S &= \frac{1}{4} \int_M d^{3}x\sqrt{|g|}\left(R -\frac{1}{a^2}(\partial\phi)^2-4V(\phi)\right)\\
    &+\fr{1}{2} \int_{\dt M} d^2 x \sqrt{\g} K + S_{B}.
    \end{aligned}
\end{equation}

In the weak AdS/CFT correspondence the above action evaluated on-shell generates one-point correlation functions in the presence of a source (the boundary data), however, generically the action \eqref{eq:action0} diverges and hence has to be regularized and renormalized. Without digging into too much details referring the reader to the original articles \cite{deHaro:2000vlm,Bianchi:2001de,Papadimitriou:2004ap,Papadimitriou:2004rz} and the reviews \cite{Skenderis:2002wp,Papadimitriou:2016yit} let us give a short overview of the general principles of the procedure. To start with one works in the Hamiltonian-Jacobi approach with the understanding that moving inside the bulk as an evolution w.r.t. the radial direction parametrized by $r$. Introducing a regularizing surface $\S_r$ diffeomorphic to the boundary $\d M = \S_{\infty}$ one notices that for a general domain wall solution not any boundary condition defined on $\S_r$ would make sense on $\dt M$. The reason is the conformal symmetry, that is in general broken on $\S_r$, and hence an arbitrary function $J_r(\phi)=0$  defining boundary conditions on $\S_r$ will break conformal invariance once the regulator is removed. The way out is to use conformal boundary conditions from the very beginning, using only local functions of fields with certain conformal weight, which are $\f_-(\vec{x})$ and $\hat{\p}_{+}(x)$. On the conformal boundary these are defined as
\begin{equation}
    \begin{aligned}
        \f_-(\vec{x}) & = \lim_{r \to \infty} e^{\D_- r} \f(r,\vec{x}), \\
         \hat{\p}_+(\vec{x}) & = \lim_{r \to \infty} e^{\D_+ r} \p_\f(r,\vec{x}), 
    \end{aligned}
\end{equation}
where $\p_\f$ is the radial canonical momentum of the field $\f(r,\vec{x})$. At arbitrary $r$ one expands the corresponding field in eigenfunctions of the dilatation operator. Dirichlet boundary conditions fix the value of $\f_-(\vec{x})$ on the boundary, Neumann boundary conditions fix the value of $\hat{\p}_+(\vec{x})$, while mixed boundary conditions imply
\begin{equation}
    \d \hat{\p}_+(\vec{x}) + f''(\f_-)\, \d \f_-\Big|_{\dt M} = 0,
\end{equation}
for an arbitrary function $f(\f_-)$. Second derivative of the function has been used for further convenience, since in this case the effective action will be deformed by the function itself.

To renormalize the action one adds local counterterms to obtain
\begin{equation}
    S_{ren} = S_{reg} + S_{ct} + S_B^{fin},
\end{equation}
where we denote by $S_{reg}$ only the regularized part of the action \eqref{eq:action0} without the $S_B$ where  $S_B^{fin}$ denotes the finite part of the boundary term. It is important to note that $S_B$ in general has an infinite part, that has to be added irrespective of the boundary conditions, and a finite part. The latter for Dirichlet boundary conditions is simply zero, while for Neumann and mixed boundary conditions reads
\begin{equation}   
    \begin{aligned}
        S_B^{fin} =& - \int d^2x \sqrt{g_{(0)}} \f_-(\vec{x}) \hat{\p}_+(\vec{x}) \\
        &+ \int d^2x \sqrt{g_{(0)}} \Big(f(\f_-) - f'(\f_-)\, \f_-\Big).
    \end{aligned}
\end{equation}
Note that Neumann b.c. are a particular case of mixed b.c. with $f(\f_-) = 0$. Using Gauss-Codazzi equations and Einstein-Hilbert equations the renormalized action in the zero-derivative approximation can be brought to the following form \cite{Martelli:2002sp}
\begin{equation}
    \label{eq:Sreg}
    S_{ren} = \int_{\S_r} d^2 x \sqrt{\g}\, W (\f) - \int_{\S_r} d^2 x \sqrt{\g}\,  U (\f)  + S_B^{fin},
\end{equation}
The function $U(\f)$ here, as was shown in \cite{Papadimitriou:2004ap}, satisfies the same equation \eqref{eq:V2W} as $W(\f)$ and necessarily has expansion \eqref{wpm1} starting with
\begin{equation}
    U(\f) = -2m\left(1 + \fr1{2a^2} \D_- \f^2 + \dots \right) \, .
\end{equation}
More generally, assuming the metric and the scalar are functions of the radial coordinate only, the effective action for the scalar field interacting with dynamical gravity in the two-derivative approximation has been obtained in \cite{Papadimitriou:2007sj} and we list the results in Table \ref{tab:actions} 
\begin{table}[http]
    \centering
    \begin{tabular}{|c|c|c|c|}
        \hline
                & D & N & M  \\
            \hline
       $J$   &  $\f_-(\vec{x})$ & $-\hat{\p}_+(\vec{x}) \displaystyle\ph{\int}$ & $-\hat{\p}_+(\vec{x}) - f'(\f_-)$\\
       \hline 
       $\s$ & $\hat{\p}_+(\vec{x})$ &  $\f_-(\vec{x}) \displaystyle\ph{\int}$ & $\f_-(\vec{x})$ \\
       \hline 
       $\Gamma$ & $I_-[-\hat{\p}_+]$ &  ${I}_+[\f_-]\displaystyle\ph{\int}$ & $I_+[\f_-] + \int_x f(\f_-)$\\[0.2cm]
       $\mc{W}$ & $\displaystyle I_+[J] $ & $\displaystyle I_-[J] \displaystyle\ph{\int}$ & $I_f[J]$ \\[0.2cm]
       \hline 
        \end{tabular}
    \caption{Holographic dictionary for a scalar field with Dirichlet, Neumann and mixed boundary conditions. Note that this is the same as Table 3 of \cite{Papadimitriou:2007sj}}
    \label{tab:actions}
\end{table}

For a 2-dimensional conformal boundary the functionals $I_\pm$ are given by \cite{Papadimitriou:2007sj}
\begin{equation}
    \begin{aligned}
        I_-[\f_+] & = \fr{1}{\D_+} \int d^2 x \sqrt{\g_0} \Big[\log \f_+ \, R[\g_0] + \fr{2}{\D_+}\f_+^{-2} (\dt\f_+)^2\Big],\\
        I_+[\f_-] & = \fr{1}{\D_-} \int d^2 x \sqrt{\g_0} \Big[\log \f_- \, R[\g_0] + \fr{2}{\D_-}\f_-^{-2} (\dt \f_-)^2\Big] \\
        &+ \int d^2 x \sqrt{\g} \Big[\x \f_-^{\fr{2}{\D_-}} + f(\f_-)\Big],
    \end{aligned}
\end{equation}
where $(\dt \f_\pm)^2 = \g^{ij}\dt_i \f_\pm \dt_j \f_\pm$ and $f(\f_-)$ is non-zero for mixed boundary conditions. These are not compatible with $W_+$-type solutions, if irrelevant deformations are not included, which is the case here. The correlation function is the same as in the case of Neumann boundary conditions while the source is given by
\begin{equation}
    J = - \hat{\p}_+(\vec{x}) - f'(\f_-).
\end{equation}
The effective action and generating function of one-point correlation functions are deformations of that of the Neumann case:
\begin{equation}
    \begin{aligned}
        \G^{mix} & = \G^{N} + \int_x f(\f_-),\\
        \mc{W}^{mix} & = \mc{W}^{N}\Big|_{f \neq 0}.
    \end{aligned}
\end{equation}
This shows that mixed boundary conditions correspond to a deformation of the theory dual to solutions with Neumann boundary conditions, as it has been shown in \cite{Papadimitriou:2007sj} and noticed earlier in \cite{Klebanov:1999tb}.

\subsection{Correlation functions}
\label{correlation}

Domain wall solutions via the AdS/CFT correspondence describe deformations of the dual CFT triggered by adding an operator or by giving a non-vanishing value to its VEV. The former breaks conformal symmetry explicitly, while the latter breaks it spontaneously. Whether a solution corresponds to either of these two cases can be determined by calculating one-point correlation functions corresponding to various boundary conditions and different types of solutions $W_\pm$. Since we are working in the zero-derivative approximation, for that one may either use the last line of Table \ref{tab:actions} dropping extra terms, or to use the renormalized action in the form
\begin{equation}
    S_{ren} = \int_{\S_r} d^2x \sqrt{\g}\Big(W(\f) - U(\f)\Big) + S_B^{fin}.
\end{equation}
Depending on the boundary conditions the source is either $\f_-(\vec{x})$, $\hat{\p}_+(\vec{x})$ or their combination. Hence, the prescription for correlation functions will be different. On top of that correlation functions certainly depend on the quantum state, that is reflected in taking either $W_+$- or $W_-$-type solutions.

Let us start with the \textbf{Dirichlet} case, where $S_B^{fin} =0$ and we have 
\begin{equation}
    \begin{aligned}
        \langle T_{ij}\rangle_{ren} & = - 2 \lim_{r \to \infty} \left[\fr{1}{\sqrt{\g}}\fr{\d S_{ren}}{\d \g^{ij}}        \right] = - 2 \hat \p_{(d)}{}_{ij} \, ,\\
        \langle \mc{O}_{\D_+}\rangle_{ren} & =  \lim_{r \to \infty} \left[e^{\D_+ r}\fr{1}{\sqrt{\g}}\fr{\d S_{ren}}{\d \f}        \right] = \hat \p_+ \, .
    \end{aligned}
\end{equation}
For the domain wall ansatz these become \cite{Martelli:2002sp}
\begin{equation}
    \begin{aligned}
        \langle T_{ij} \rangle_{ren} & = \lim_{r\to \infty} \left[\hat{g}_{ij}(W(\f) - U(\f))\right], \\
        \langle \mc{O}_{\D_+} \rangle_{ren} & =- \lim_{r\to \infty} \left[e^{\D_+ r}(W'(\f) - U'(\f))\right],
    \end{aligned}
\end{equation}
Now, for $W_+$ type solutions we have the following
\begin{equation}
    \begin{aligned}
        \langle T_{ij} \rangle_{ren} & = \lim_{r\to \infty} \left[\hat{g}_{ij}(W_+(\f) - U(\f))\right] \\
        & = -\fr12 2 \n\lim_{r \to \infty} \left[\hat{g}_{ij}e^{-2\D_+ r}e^{2 r}  \hat{\f}_+^2 \right] = 0,\\
        \langle \mc{O}_{\D_+} \rangle_{ren} & =- \lim_{r\to \infty} \left[e^{\D_+ r}(W_+'(\f) - U'(\f))\right] \\
        & =(2 - 2\D_+) \hat \f_+(\vec{x}).
    \end{aligned}
\end{equation}
Here we denote $\D_\pm = 1 \pm \n$ and take into account the near horizon behavior of the scalar field and the metric 
\begin{equation}
    \begin{aligned}
        \phi(r,\vec{x}) & \sim e^{- \D_+ r}\hat{\f}_+(\vec x) + e^{- \D_- r}\hat{\f}_-(\vec x), \\
        g_{ij}(r,\vec{x}) & \sim e^{2 r} \hat{g}_{ij}.
    \end{aligned}
\end{equation}
For $W_-$ type solutions we simply have zero for both correlation functions since $U= W_-$ have
\begin{equation}
    \begin{aligned}
         \langle T_{ij} \rangle_{ren} & =  0,\\
        \langle \mc{O}_{\D_+} \rangle_{ren} & = 0.
    \end{aligned}
\end{equation}
Note also that the action $S_{ren}$ vanishes for both cases once the regulator  $r \to \infty$, i.e. when the UV cut-off is removed.

We see, that RG flows described by $W_+$ type solutions are triggered by a non-vanishing expectation value of a (relevant) scalar operator of conformal dimension $\D_+$, while the action is not changed. From the third row of Table \ref{tab:actions} it is easy to see, that $W_-$ type solution correspond to a single-trace deformation of the initial theory sitting at the fixed point. Alternatively, one can show that as follows. First, since $\D_- < \D_+$ a $W_+$ type solution must behave near the boundary as
\begin{equation}
    \f_+(r,\vec{x}) \sim e^{-\D_+ r}\hat{\f}_+(\vec{x}),
\end{equation}
i.e. must have no contribution of weight $\D_-$. In the CFT language this corresponds to having vanishing source on the boundary. On the contrary, a $W_-$ type solution is necessarily  of the form
\begin{equation}
    \f_-(r,\vec{x}) \sim e^{-\D_+ r} \f_+(\vec{x}) +  e^{-\D_- r} \f_-(\vec{x}),
\end{equation}
i.e. having a non-vanishing source contribution. In the CFT language this corresponds to adding a term to the boundary theory of the form
\begin{equation}
    \D S = - \int \f_-(\vec{x}) \mc{O}_{\D_+}(\vec{x}) d^2 x,
\end{equation}
that is precisely a single-trace deformation. Note, that although $\f_+(\vec{x})$ is non-zero, the renormalized correlator vanishes. This is due to the supersymmetric choice of the renormalization scheme, for which the on-shell action is zero.

\begin{table}[http]
    \centering 
    \begin{tabular}{|c|c|}
    \hline
         b.c & $W_+$    \\
    \hline
           D & \parbox{7cm}{\centering
           $\langle T^{ij} \rangle  =0 $ \\
           $\langle \mc{O}_{\D_+} \rangle =(2-2 \D_+) \hat{\f}_+(\vec{x})$\\
           source $J(\vec{x}) = \f_-(\vec{x})$, \\
           RG flow triggered by VEV\\
           
           }
           \\[0.2cm]
    \hline
           N & \parbox{7cm}{\centering 
           $\langle T^{ij} \rangle  =0$ \\
           $\langle\mc{O}_{\D_-}\rangle = 0 $ \\
           source $J(\vec{x}) = -\hat{\p}_+(\vec{x})$ \\
           RG flow by a single-trace deformation\\
           $\displaystyle S_{eff} = -\int_x J(\vec{x}) \mc{O}_{\D_- }(\vec{x}) $ \\   
           } 
           \\
           \hline
           M & \parbox{7cm}{\centering 
           incompatible if irrelevant deformations are not allowed
           } 
           \\[0.2cm]
    \hline
    \end{tabular}\\[0.2cm]
    \begin{tabular}{|c|c|}
    \hline
         b.c & $W_-$   \\
    \hline
           D & 
           \parbox{7cm}{\centering
           $\langle T^{ij} \rangle  =0$ \\
           $\langle \mc{O}_{\D_+} \rangle = 0$ \\
           source $J(\vec{x}) = \f_-(\vec{x})$, \\
           RG flow by a single-trace deformation \\
           $\displaystyle S_{eff} = -\int_x J(\vec{x}) \mc{O}_{\D_+}(\vec{x}) $ \\ 
           }  
           \\[0.2cm]
    \hline
           N  & 
           \parbox{7cm}{\centering
           $\langle T^{ij} \rangle  =0$ \\
           $\langle \mc{O}_{\D_-} \rangle = \f_-(\vec{x})$\\
           source $J(\vec{x}) =  -\hat{\p}_+(\vec{x})$, \\
           RG flow triggered by VEV\\
           }
           \\
           \hline
           M& 
           \parbox{7cm}{\centering
           $\langle T^{ij} \rangle  = - f(\f_-)g_{(0)}{}^{ ij}$ \\
           $\langle \mc{O}_{\D_-} \rangle = \f_-(\vec{x})$ \\
           $S_{eff} = S_{N} + \int_x \sqrt{\g} f(\f_-)$ \\
           A multitrace deformation of the theory corresponding to Neumann b.c. 
           }
           \\
    \hline
    \end{tabular}
    \caption{Holographic dictionary between solutions of the types $W_\pm$ with various boundary conditions (Dirichlet, Neumann and mixed)  and RG flows triggered by a deformation or a non-vanishing VEV. Here only relevant and unitary operators of scale dimension $1 <  \D_+ < 2$, are considered.}
    \label{tab:interpretations}
\end{table}

For \textbf{Neumann} boundary conditions the source is given by the component $-\p_+(\vec{x})$ of the radial momentum $\p_\f$ and hence for the scalar operator, which is now of conformal weight $\D_-$, we have
\begin{equation}
    \begin{aligned}
        \langle \mc{O}_{\D_-}\rangle_{ren} & =  \lim_{r \to \infty} \left[e^{\D_- r}\fr{1}{\sqrt{\g}}\fr{\d S_{ren}}{-\d \p_\f}        \right],
    \end{aligned}
\end{equation}
where here we understand $S_{ren}$ as a functional of the momentum as in \eqref{eq:corrapp} (including finite terms). Apparently, nothing changes for the energy momentum tensor, as boundary conditions for the metric are always Dirichlet. Now the on-shell action differs from that for the Dirichlet case by the finite part of the boundary term and the function $U(\f)$ is still the same. However, since in the Hamiltonian approach momentum and canonical variable are independent, what varies is only the boundary term, that gives
\begin{equation}
    \begin{aligned}
        & W_+ : && \langle \mc{O}_{\D_-}\rangle_{ren} = 0, \\
        & W_- : && \langle \mc{O}_{\D_-}\rangle_{ren} = \f_-(\vec{x}).
\end{aligned}
\end{equation}
Now the interpretation is that solutions corresponding to $W_+$ are holographically dual to RG flows triggered by adding a relevant operator, while solutions corresponding to $W_-$ are dual to theory in a phase with conformal symmetry spontaneously broken by a VEV.

Finally, for \textbf{mixed} boundary conditions the finite part of the boundary action is 
\begin{equation}
    S_B^{fin} = \int_x \big(\f_-(\vec{x})J(\vec{x}) +f(\f_-) \big) \, ,
\end{equation}
where $J(\vec{x}) = -\hat{\p}_+(\vec{x}) - f'(\f_-)$, and hence the correlation function is that same as in the Neumann case. However, the dual CFT interpretation is now different: solutions with mixed boundary conditions correspond to a multitrace deformation of that with Neumann boundary conditions. We collect the results reviewed in the above  discussion in Table \ref{tab:interpretations}. Before we continue with the RG flow analysis let us note that the way supersymmetry works at the conformal boundary of AdS with different boundary conditions was investigated in \cite{Deger:2000as}.

\section{Domain wall solutions}
\label{sec:sols}

\begin{figure}[H]
    \centering
        \begin{subfigure}{0.45\textwidth}
            \centering
             \includegraphics[width = \textwidth]{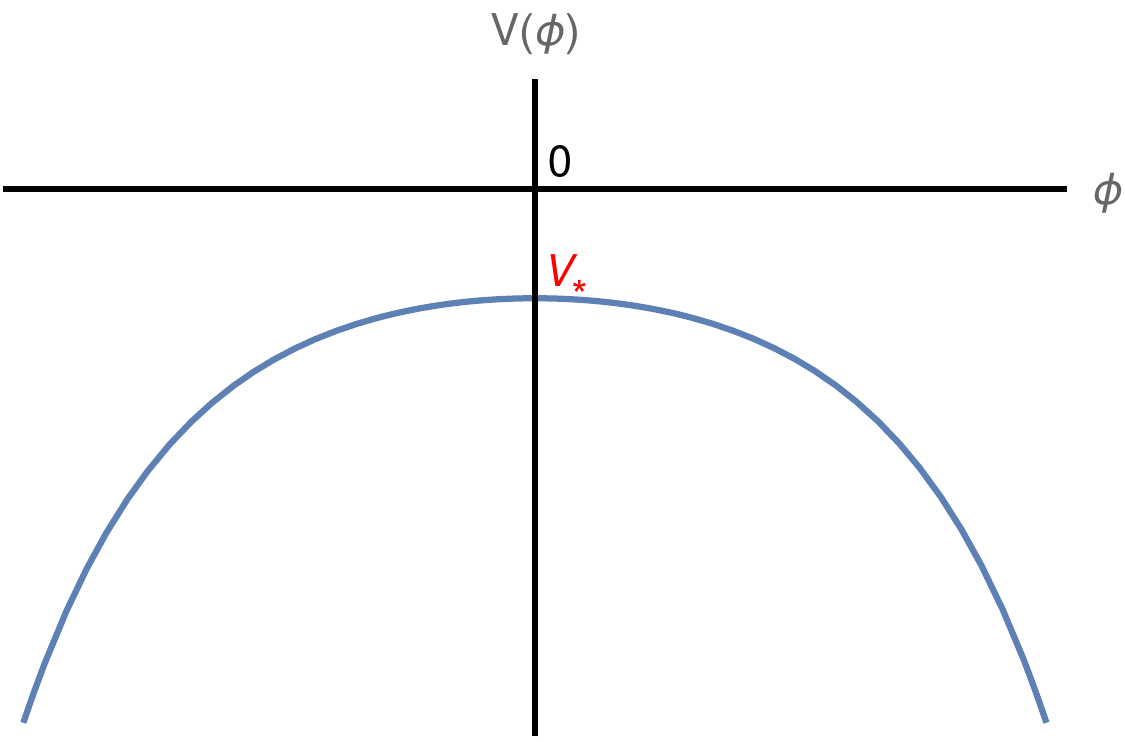}
             \caption{ $0 <a^2 \leq \fr12$}
        \end{subfigure}
        \hfill
        
        \begin{subfigure}{0.45\textwidth}
            \centering
             \includegraphics[width = \textwidth]{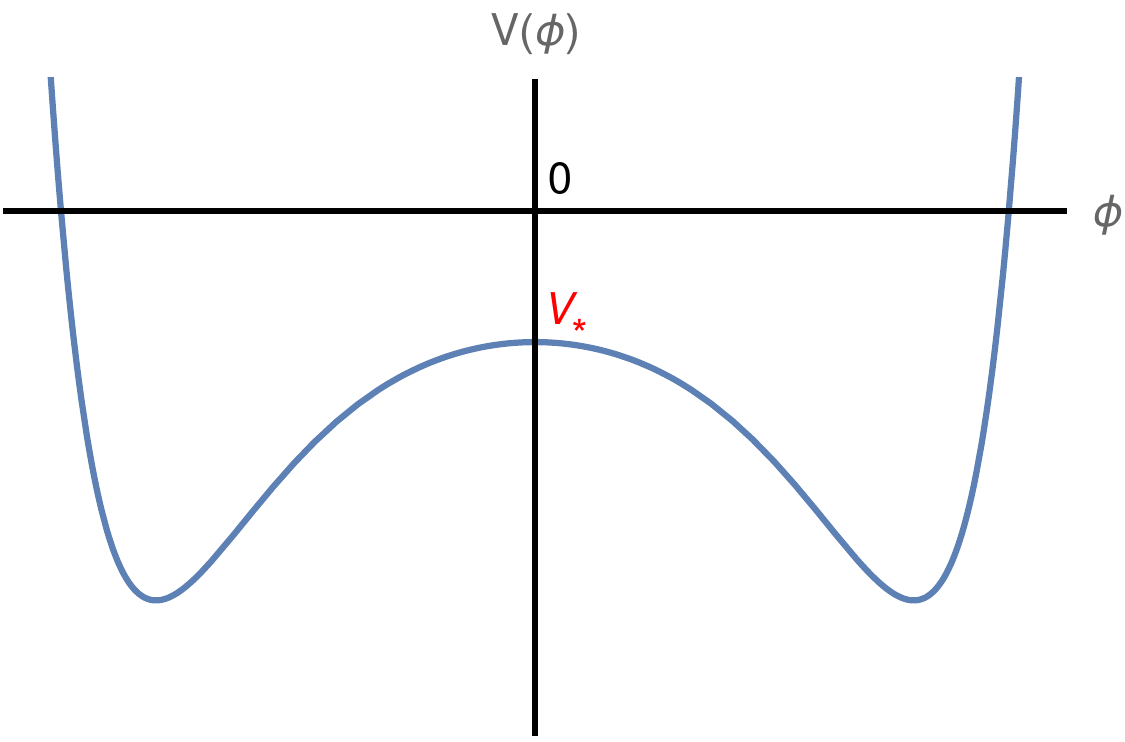}
             \caption{$\fr12 < a^2 < 1$}
        \end{subfigure}    \hfill     
        \begin{subfigure}{0.45\textwidth}
            \centering
             \includegraphics[width = \textwidth]{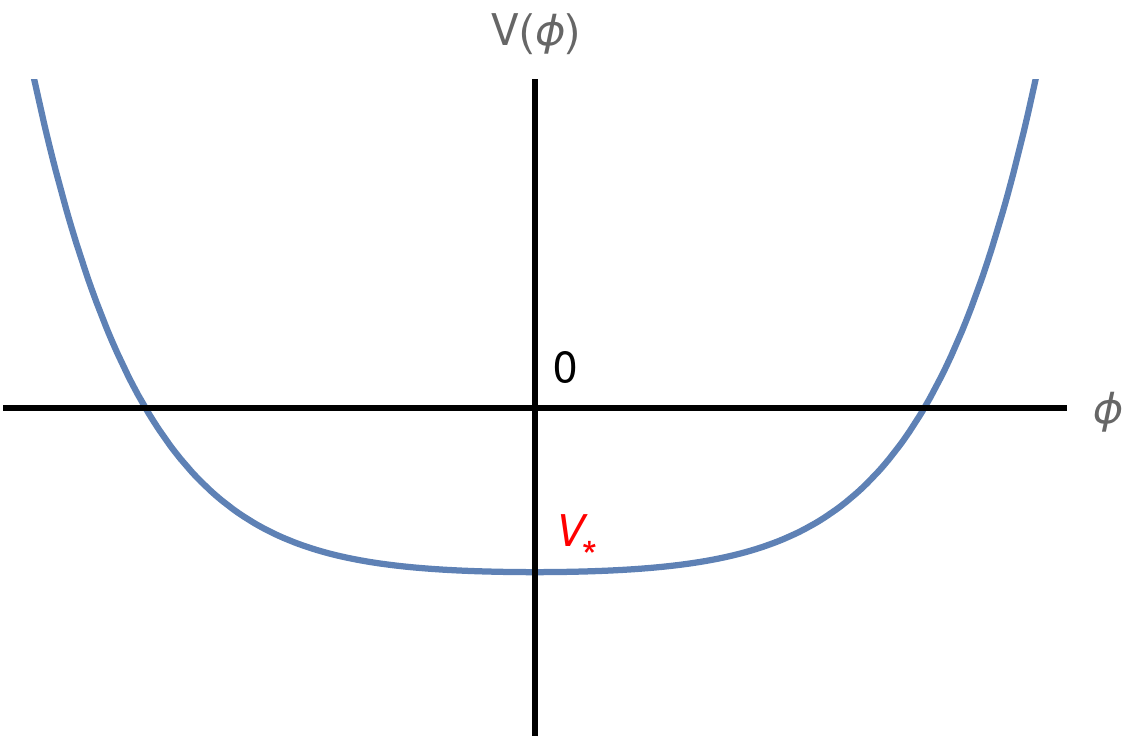}
             \caption{$a^2 \geq 1$}
        \end{subfigure}
    \caption{The scalar potential for different values of $a^2$ where $V(0)=-2m^2$.}
    \label{fig:Vplot}
\end{figure}

We now start looking at domain wall solutions \eqref{metricT} of  our model \eqref{eq:sugra} in detail. Its scalar potential \eqref{pot} can be rewritten in terms of the field $\phi$ as
\begin{equation}
\label{eq:potential}
V(\phi)=- 2{m^2}\cosh^2\phi\left[(1-2a^2)\cosh^2\phi+2a^2\right].
\end{equation}
Depending on the value of the non-zero constant $a^2$ the potential has qualitatively different behavior and, as we will see later, the phase picture of RG flows differs crucially. Plots of the potential $V(\phi)$ are demonstrated on Fig. \ref{fig:Vplot}.

For our model there exists an analytic half-superymmetric domain wall solution of the from \eqref{metricT} to the field equations valid for any $a^2$, for which the scale factor is given by \cite{Deger:2002hv}
\begin{equation}\label{scafDeg}
	A _{\rm susy} = \frac{1}{4a^{2}}\ln (e^{8 m a^2 r}-1),
	\end{equation}
so the metric reads
\begin{equation}\label{metricDeg}
ds^2_{\rm susy}=(e^{8 m a^2 r}-1)^{\frac{1}{2a^2}}\left(-dt^2+dx^2\right)+dr^2.
\end{equation}
The scalar field supporting the solution has the following profile 
\begin{equation}\label{scafieldDeg}
	\phi_{\rm susy} = \frac{1}{2}\ln\left(\frac{1+e^{-4ma^2 r}}{1-e^{-4m a^2 r}}\right), \quad 0\leq r< \infty \, .
 \end{equation}
As we have discussed before, the superpotential corresponding to this solution is given by\footnote{Note the minus sign comparing to \cite{Deger:2002hv}, that is related to a different choice of the direction of $r$.}
\begin{equation}
\label{superpotential}
W_{\rm susy} = -2 {m} \cosh^{2}\phi,
\end{equation}
which is the actual superpotential of the underlying $D=3$, $\mc{N}=(2,0)$ gauged supergravity coupled to a sigma model with target space $\HH^2$ \cite{Deger:2002hv}. This is as expected since the solution given above preserves half of the supersymmetry of the model \cite{Deger:1999st}. Supersymmetric solutions of this model including generalizations of the above domain wall solution were studied in \cite{Deger:2004mw,Deger:2006uc}. Note, that for the RG flow \eqref{metricDeg},\eqref{scafieldDeg} the Dirichlet boundary condition was used, which respects  supersymmetry requiring $\langle T_{ij}\rangle = 0$ \cite{Bianchi:2001de}, and hence the RG flow is supersymmetric.

We are interested in solutions that interpolate between critical points of the potential (for $1/2 < a^2 < 1$) or between a critical point and the minus infinity (where the weak gauge/gravity correspondence holds). The potential \eqref{eq:potential} has at least 1 and at  most 3 critical points
\begin{equation}
    \begin{aligned}
        \f_{* 1} & = 0,  \\
        \f_{* 2,3} & = \fr12 \log\left[\fr{1\pm 2\sqrt{a^2 - a^4}}{2a^2 -1}\right], \quad \mbox{for } \quad \fr12 < a^2  < 1.
    \end{aligned}
\label{critical}
\end{equation}
The first one is supersymmetric and it exists for all $a^2$ whereas the other two are not supersymmetric and exist only when $\fr12 < a^2  < 1$ \cite{Deger:1999st}.
For each critical point $\phi_*$  we now check whether the Breitenlohner-Freedman-(BF) bound \cite{Breitenlohner:1982jf} is satisfied. First notice that for all three critical points 
\eqref{critical} the domain wall metric \eqref{metricT} is a solution with constant scalar field and describes AdS geometry.
To see this, note that by setting $\phi(r)=\phi_*$ the field equation \eqref{eqd} is satisfied identically. Then, the remaining field equations \eqref{eom1} and \eqref{eom3} 
are solved by 
\begin{equation}
    A(r) = \sqrt{-2V_*} r,
\end{equation}
where $V_*=V(\f_*)<0$ is the value of the potential at a critical point. From this we see that the metric \eqref{metricT} is nothing but the AdS${}_3$ metric in the Poincar\'e patch.

To determine the BF bound imposed by stability of a scalar field in an AdS space-time we expand the potential \eqref{eq:potential} near the critical points \eqref{critical} up to the quadratic order 
\begin{equation}
    \begin{aligned}
         *1) && V(\f) & \simeq - 2 {m^2} + 4m^2(a^2 - 1)\f^2, \\
         *2,3) && V(\f) & \simeq \fr{2a^4 {m^2}}{2a^2 -1} - \fr{8a^2(a^2-1){m^2}}{2a^2 -1} (\f-\f_{*2,3})^2.
    \end{aligned}
\end{equation}
Now, from the scalar field equation \eqref{scalar}
we read the effective scalar field masses at critical points as
\begin{equation}
    \begin{aligned}
        1) && M_1^2 =2a^2 V''(\phi_{*1})=   16 {m^2} a^2(a^2 - 1), \\
        2,3) && M_{2,3}^2= 2a^2 V''(\phi_{*2,3})=   -   \fr{32{m^2}a^4(a^2-1)}{2a^2 -1} \, .
    \end{aligned}
    \label{BF}
\end{equation}
The BF bound for AdS$_3$ with radius $1/2m$ requires $M^2 \geq -{4m^2}$, and from the above it is easy to see that all critical points of the model satisfy this condition.

For the domain wall solutions interpolating between a critical point $\phi_*$ and a different vacuum at the region $r \to \infty$ brings us to a critical point $\f_*$ in the field space. Hence, near a critical point one may consider the solution to be that of the near horizon region of the AdS$_3$ space, meaning that the scalar field behaves as 
\begin{equation}\label{Sc-an}
\phi\sim e^{-{2m}\,\Delta\,  r},
\end{equation}
where the parameter $\Delta$ corresponds to the scaling dimension of the dual operator. {In what follows we set the AdS$_3$ radius to 1, i.e. $\frac{1}{2m}=1$, which fixes the cosmological constant of the AdS$_3$ to $\Lambda = -1$}. The AdS/CFT duality relates the mass of the scalar field and the conformal weight $\Delta$ of the corresponding boundary operator as
\begin{equation}\label{BFound2}
\Delta(\Delta- 2)-M^2=0 ,
\end{equation}
from which we recover two branches of the scaling dimension in terms of the parameter $a^2$ at different critical points:
\begin{equation} \label{dimension}
    \begin{aligned}
        &\f_{*1}:&& \Delta_{\pm}=\left(1\pm|1- 2a^2|\right), \\
        &\f_{*2,3}:&& \D_{\pm} =\left( 1 \pm  \sqrt{1 + \fr{8a^4(1-a^2)}{2a^2 -1 }}\right).
    \end{aligned}
\end{equation}
 It is worth noting that since the dual conformal field theory is two-dimensional, the operators are relevant for $\Delta<2$, marginal for  $\Delta=2$ and irrelevant for  $\Delta>2$. The unitarity bound is $\D \geq 0$.

For operators of the theory at the critical point $\f_{*1}$ we have the following cases for the scaling dimensions $\D_\pm$ depending on $a^2$ \cite{Deger:2002hv}:
\begin{enumerate}
    \item For $0<a^2 < 1$: $1 \leq \D_+ < 2$ and $0 < \D_- \leq 1$ , hence RG flow of both dual CFT's are given by a relevant operator.
    \item For $a^2 >1$: $\D_+ > 2$  corresponding to an irrelevant operator and $\D_- <0$ which violates the unitarity bound.
    \item For $a^2 = 1$: $\D_+ = 2$ corresponds to a marginal operator and $\D_- = 0$ saturates the unitarity bound.
\end{enumerate}
Hence, we conclude that for the critical point $\f_{*1}$ to describe an RG flow triggered by adding a relevant operator to its action or by its VEV we must stay in the region $0 < a^2 < 1$. On the other hand, for the critical points $\f_{*2,3}$ we have  $1/2< a^2 < 1$ which implies $\D_+ > 2$ corresponding to an irrelevant operator, and $\D_- < 0$ which breaks the unitarity bound.

As we have discussed in Section \ref{correlation} a generic profile for the scalar field near the conformal boundary reads:
\begin{equation}\label{ScalarFieldGen} 
\phi \sim 
\phi_{-}e^{- \D_- r}+ \phi_{+} e^{-\Delta_+ r},
\end{equation}
and is of the $W_+$ or $W_-$ type depending on whether $\f_-$ is zero or not respectively (since $\D_+ > \D_-$). Depending on boundary conditions these have different interpretations on the field theory side (see Table. \ref{tab:interpretations}).

\subsection{Fake superpotential near critical points}

Recall that fake superpotential $W$ is defined by the relation 
\begin{equation} \label{VthroughW}
V(\phi) = \frac{a^2}{4}\left(\frac{\partial W}{\partial \phi}\right)^2 -\frac{1}{2}W^2,
\end{equation}
and is considered as a solution to this differential equation. In general, there exists a family of such solutions of which the actual superpotential $W_{\rm susy}$ is a representative. However, usually it is a very nontrivial task to solve such a differential equation analytically, and hence approximate methods and near boundary analysis become very important.

From the discussion around \eqref{wpm1}, we know that near a critical point $\f_*$ the superpotential can be expanded in two ways as (with $2m=1)$:
\begin{equation} \label{wpm}
W_{\pm}(\phi)  = -\sqrt{-2V_*} - \frac{1}{2a^2}\D_\pm(\phi - \f_*)^2 + \ldots ,
\end{equation}
where we have chosen $W(\f_*)< 0$ to describe supersymmetric flows. Note that $W'_{\pm}(\phi_*)=0$, that is $\phi_*$ is a critical point of $W_\pm$ as well.
Here, values of the scaling dimensions in the expression above must be those corresponding to the chosen critical point. 

Although solving the quadratic equation \eqref{VthroughW} leads to two branches of the fake superpotential, the actual behavior is determined by the solution to field theory equations and is not a matter of choice. Hence, all solutions can be divided into two groups: of $W_+$ and of $W_-$ type. Let us illustrate this on the analytical solution \eqref{scafDeg}-\eqref{superpotential}
studied in \cite{Deger:2002hv}, that describes an RG flow starting at the critical point $\f_{*1}$. From \eqref{dimension} we see that the scaling dimensions are $\D_\pm = 1 \pm |1-2a^2|$ in this case  and from \eqref{wpm} with $V_*=-1/2$ and $\f_{*1}=0$ the branches of the fake superpotential become
\begin{equation} \label{wexpansion}
    \begin{aligned}
        &2a^2 > 1: && \left\{
            \begin{aligned}
                W_+ & = -1 - \f^2 + \mc{O}(\f^3), \\
                W_- & = -1 - \frac{(1-a^2)}{a^2} \f^2+ \mc{O}(\f^3) ,                 
            \end{aligned}\right. \\
        &2a^2 = 1: && W_\pm = -1 - \f^2+ \mc{O}(\f^3) , \\
        &2a^2 < 1: && \left\{
            \begin{aligned}
                W_+ & = -1 - \frac{(1-a^2)}{a^2} \f^2 + \mc{O}(\f^3), \\
                W_- & = -1 - \f^2+ \mc{O}(\f^3) ,                 
            \end{aligned}\right.  
    \end{aligned}
\end{equation}
Let us now compare the above behavior with the exact superpotential \eqref{superpotential}. 
Clearly its expansion near $\f_*1$ does not depend on $a^2$ and is always of the form
\begin{equation}
    W_{\rm susy} = -1 - \f^{2} + \mc{O}(\f^3).
\end{equation}
Hence, we conclude that
\begin{equation} \label{Wpm}
    \begin{aligned}
        & 2a^2 > 1: && W_{\rm susy} = W_+, \\
        & 2a^2 < 1: && W_{\rm susy} = W_-.
    \end{aligned}
\end{equation}
The value $2a^2 =1$ is distinguished since $W_\pm$ are the same and equal to $W_{\rm susy}$ and therefore needs a special treatment. Using Table \ref{tab:interpretations} one can easily determine the  RG flow dual to the solution depending on the value of $a^2$ and boundary conditions. For the supersymmetric solution \eqref{scafDeg}-\eqref{superpotential}
analyzed in \cite{Deger:2002hv} the boundary behavior of the scalar field is fixed and corresponds to a Dirichlet type boundary condition case. As was found in \cite{Deger:2002hv}, the RG flow is generated by a VEV when $a^2>1/2$ and by a deformation for $a^2 < 1/2$ which is in agreement with \eqref{Wpm} and the Table \ref{tab:interpretations}. In the next section we will reach the same conclusions using the dynamical systems approach. Moreover, we will look at the special case $a^2=1/2$ in detail and see that the supersymmetric flow is triggered by a VEV.

\subsection{Holographic RG flows from dynamical system equations}

As we have already mentioned the solution \eqref{scafDeg}-\eqref{scafieldDeg}  of \cite{Deger:2002hv} with the superpotential \eqref{superpotential} is only a representative of a set of domain wall solutions to the differential equation \eqref{VthroughW}. Following the discussion in \cite{Papadimitriou:2007sj} we conclude, that for $2a^2 > 1$ the corresponding superpotential can be obtained as a limit of a family of solutions $W_\h=W(\f,\h)$ as the parameter $\h$ tends, say, to infinity. On the other hand, for $2a^2 \leq 1$ the superpotential \eqref{VthroughW} must be a representative of some (possibly different) family of solutions $\tilde{W}(\f,\q)$ corresponding to a fixed value of the parameter $\q$. For the potential \eqref{eq:potential} the differential equations become too complicated for us to find solutions analytically, leaving an approximate analysis as the only option. An example of such a solution was given in \cite{Deger:2002hv} for $1/2 < a^2 < 1$ , where a domain wall interpolating between the supersymmetric AdS critical point $\f_{*1}$ and  the non-supersymmetric AdS critical points $\f_{*2}$ or $\f_{*3}$ together with the corresponding fake superpotential were found numerically.

Here we follow the dynamical system approach based on the representation of domain wall solutions \eqref{metricT} of the supergravity field equations in terms of mechanical equations of a dynamical system evolving along the energy scale $A=A(r)$. Variables of the dynamical system are chosen to be as follows
\begin{equation}
    \begin{aligned}
        Z & = \fr{1}{1+e^{\phi}} \implies \quad \phi = \ln\Big(\frac{1-Z}{Z}\Big), \\
        X & = \frac{\dot{\phi}}{\dot{A}}. 
    \end{aligned}
\end{equation}
Note, that $Z$ tends to $0$ as $\phi \to \infty$ and goes to $1$ as $\phi\to-\infty$, at $\f=0$  $Z=1/2$. Hence, this choice allows to represent the RG flow  to infinitely large values of $|\f|$ on a compact plot. 

The first order field equations \eqref{first} then become
(see Appendix \ref{app:autonom} for the derivation): 
\begin{equation}
\label{eqZX}
    \begin{aligned}
        \frac{dZ}{dA} & =   X Z(Z-1) , \\
        \frac{dX}{dA} & = \bigg(\frac{X^2}{a^2} - 2 \bigg)\bigg(X + \frac{a^2}{2} \fr{V'}{V} \bigg). 
    \end{aligned}
\end{equation}
When the RHS of both equations vanish there is a critical point of the system which we denote as $p_\a = (Z_\a,X_\a)$. Near these points behavior of the system can be investigated using the Lyapunov analysis of the dynamical systems. For a two-dimensional system as ours one has the following equations
\begin{equation}
    \fr{d}{dA} X^i = F^i(Z, X),
\end{equation}
where $i=1,2$. Near a critical point $(X^1_0,X^2_0)$ the RHS can be expanded up to the linear order to give
\begin{equation}
    \fr{d}{dA} X^i = \cM^i{}_j (X^j - X_0^j),
\end{equation}
where
\begin{equation}
    \cM^i{}_j = \fr{\dt F^i}{\dt X^j}\Bigg|_{X_0}.
\end{equation}
The obtained linear equation can be integrated and we arrive at the following general solution to the initial set of equations near a critical point
\begin{equation}
    X^i = X_0^i + e^{\lambda_1 A} u^i_1 + e^{\lambda_2 A} u^i_2,
\end{equation}
where $u_{1,2}$ are eigenvectors of the matrix $\cM^i{}_j$ corresponding to the eigenvalues $\lambda_{1,2}$:
\begin{equation}
    \cM^i{}_j u_{1,2}^j = \lambda_{1,2} u^i_{1,2}.
\end{equation}
The cases when $\lambda_{1,2} = 0$ or $\lambda_{1} = \lambda_{2}$ must be treated separately. For RG flows described by domain wall solutions such eigenvalues $\lambda_{1,2}$ will be nothing but $\D_\pm$ and hence will distinguish between $W_\pm$ types of the solution. Specifying boundary conditions we will be able to identify the field theory picture for each phase trajectory.

Let us now return to the equations \eqref{eqZX} with the potential \eqref{eq:potential}. This system possesses several critical points three of which with $X=0$ correspond to AdS critical points of the potential. The full list of critical points with coordinates $(Z,X)$ is as follows:
\begin{equation}
    \begin{aligned}
        p_1 & = (1/2,0), \\
        p_{2,3} & = \left(\frac{1}{2}\pm\frac{1}{2} \sqrt{\frac{-3 a^2+2 \sqrt{a^2 \left(2 a^2-1\right)}+1}{a^2-1}},0\right),\\
        p_4 & = (0, -2a^2), \quad p_{5,6} = (0, \pm \sqrt{2}a),\\
        p_7 & = (1, -2a^2), \quad p_{8,9} = (1, \pm \sqrt{2}a), \\
        q_2 & = (0, -1/2), \quad q_3 = (1, 1/2) \, .
    \end{aligned}
\end{equation}
Here $p_1$ is the supersymmetric AdS vacuum $\phi_{*1}$ and the other 2 non-supersymmetric AdS vacua $\phi_{*2,3}$ corresponding to points $p_{2,3}$ exist only when $1/2< a^2 < 1$. The points $p_{4,7}$ do not exist for $a^2=1/2$. Finally, $q_{2,3}$ exist only for $a^2=1/2$.
Types of these critical points are listed in Table \ref{tab:types}.

Critical points with $Z=0,1$, i.e. $p_{4,7}$, $q_{2,3}$ correspond to the scalar potential $V\to \pm \infty$ with $\phi\to \pm \infty$. The geometries near these points, to which gravity solutions asymptote, are scale-invariant, but don’t possess conformal invariance. Moreover, they are singular since the divergence of the scalar field leads to the divergence of curvature invariants of the metric. However, the singularities can be acceptable, if they obey the so called Gubser’s criterion \cite{Gubser:2000nd} and/or the condition for the spectral computability. Then, ”good” singular solutions are not worthless in holography and can be interpreted as IR/UV fixed points. Note, that holographic RG flows with such fixed points were classified in \cite{Kiritsis:2016kog}. Following the analysis from \cite{Kiritsis:2016kog} one conclude that the solutions, which are related with points $p_{4,7}$, have acceptable singularities only for $a^2< \frac{1}{2}$, since only for this case the scalar potential grows slower than $\exp{\left(2\frac{\sqrt{2}}{a}\phi\right)}$ as $\phi \to \infty$, i.e. they satisfy Gubser's bound\footnote{ The multiplier $\frac{\sqrt{2}}{a}$ in $\exp{\left(2\frac{\sqrt{2}}{a}\phi\right)}$ for the estimation is found from eq.\eqref{eq:V2W}.}. For the same reason the geometries corresponding to the critical points $q_{1,2}$ can be classified as "good" singular solutions.

\begin{table}[ht]
    \centering
    \begin{tabular}{|c||c|c|c|c|}
    \hline
                &  $0<a^2 < \fr12$ & $a^2 = \fr12$ & $\fr12 < a^2 < 1$ & $a^2 \geq 1$ \\
    \hline
    \hline
         $p_1$  &  \multicolumn{3}{c|}{unstable node} & saddle \\
         \hline
       $p_{2,3}$&  ---     &  --- &saddle & --- \\
       \hline
       $p_{4,7}$    & saddle & --- & \multicolumn{2}{c|}{stable node} \\
       \hline
       $p_{5,9}$    & \multicolumn{2}{c|}{stable node} & \multicolumn{2}{c|}{saddle} \\
       \hline
       $p_{6,8}$    & \multicolumn{4}{c|}{saddle}\\
       \hline 
       $q_{2,3}$    &  --- & saddle & --- & --- \\
       \hline
    \end{tabular}
    \caption{Classification of critical points of the dynamical system. Long dash ``---'' means that the critical point does not exist for the given value of $a^2$. The point $p_1$ for $2a^2=1$ is a degenerate unstable node.  }
    \label{tab:types}
\end{table}
Before turning to a general analysis of RG flows, let us look at the solution given by \eqref{scafDeg},\eqref{metricDeg} from the point of view of the Lyapunov analysis. The corresponding flow starts at the point $p_1=(1/2,0)$, which is either an unstable node or a saddle. Analytically we see, that the flow ends at the point $p_4$, which will also be seen at the phase diagram. Eigenvectors and eigenvalues at $p_1$ can be found to be
\begin{equation}
    \begin{aligned}
        \lambda_1 & = -2 a^2, && \lambda_2 = -2(1-a^2)\\
        u_1 & = 
         \begin{bmatrix}
            1 \\
            -\lambda_1
         \end{bmatrix}, &&
         u_2 = 
         \begin{bmatrix}
            1 \\
            -\lambda_2
         \end{bmatrix}.
    \end{aligned}
\end{equation}
The critical point $(Z_1,X_1)=(1/2,0)$ of the dynamical system corresponds to the critical point $\f_{1}=0$ of the potential, that implies $A(r) = 2 r$. Using the values $\D_\pm = 1 \pm |2a^2 -1|$ found in the previous section we have:
\begin{equation}
    \begin{aligned}
       & 2a^2 < 1: && \lambda_1 = -\D_-, && \lambda_2 = - \D_+; \\
       & 2a^2 > 1: && \lambda_1 = -\D_+, && \lambda_2 = - \D_-.    
    \end{aligned}
\end{equation}
From the true superpotential \eqref{superpotential} found in \cite{Deger:2002hv} we have concluded in the previous Section that for $2a^2>1$ the flow is described by a $W_+$-type solution, while for $2a^2<1$ the solution is a $W_-$-type \eqref{Wpm}. Let us now consider these cases separately and show that one arrives at the same conclusions from the phase pictures too.

\subsubsection{\texorpdfstring{$0<a^2<1/2$}{a2<1/2}}

\begin{figure}[ht]
    \centering
    \includegraphics[height=8.2cm]{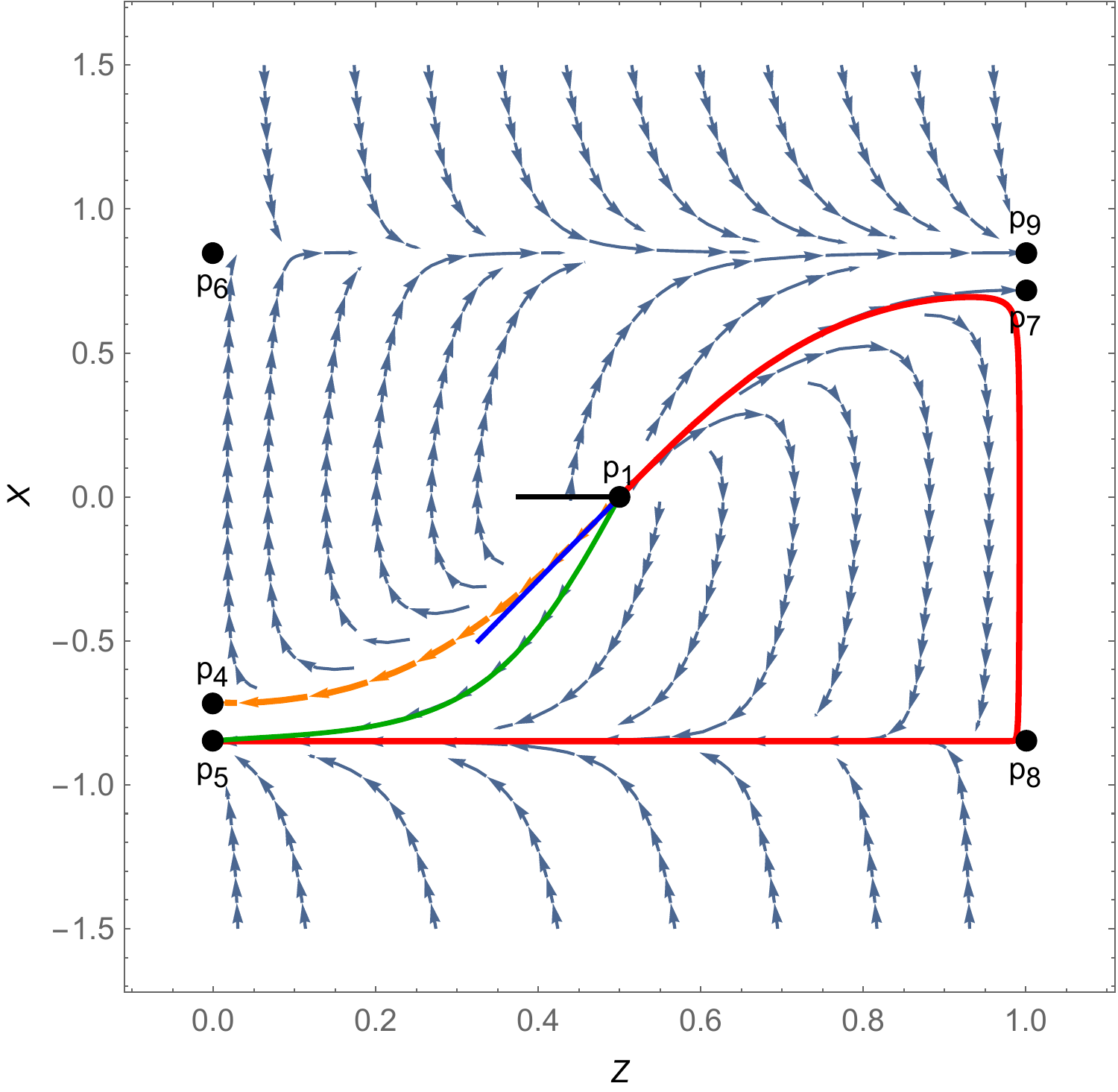}
    \caption{Phase flow for $0<a^2 <1/2$ ($a^2 = 0.36$) on the $(Z,X)$-plane. The orange line between $p_1$ and $p_4$ depicts the supersymmetric RG flow found in \cite{Deger:2002hv}. Red line denotes the flow starting at $p_1$ ending at $p_5$ at $Z=0$. The blue  and black lines denote the directions of the eigenvectors $u_1$ and $u_2$ respectively. The latter starts the $W_+$-type solution.}
    \label{fig:am12}
\end{figure}

In the region $0< 2a^2 < 1$ the general solution of the flow equations \eqref{eqZX} reads
\begin{equation}
    \begin{bmatrix}
        Z\\
        X
    \end{bmatrix}
    =
    \begin{bmatrix}
        1/2\\
        0
    \end{bmatrix}
    + k_1 e^{- \D_- A} u_1 + k_2 e^{-\D_+ A} u_2,
\end{equation}
with vectors $u_{1,2}$ given by
\begin{equation}
    \begin{aligned}
        &u_1 = 
            \begin{bmatrix}
                -1\\
                -8a^2
            \end{bmatrix}, &&
        u_2 = 
            \begin{bmatrix}
                -1\\
                -8(1-a^2)
            \end{bmatrix}.
    \end{aligned}
\end{equation}

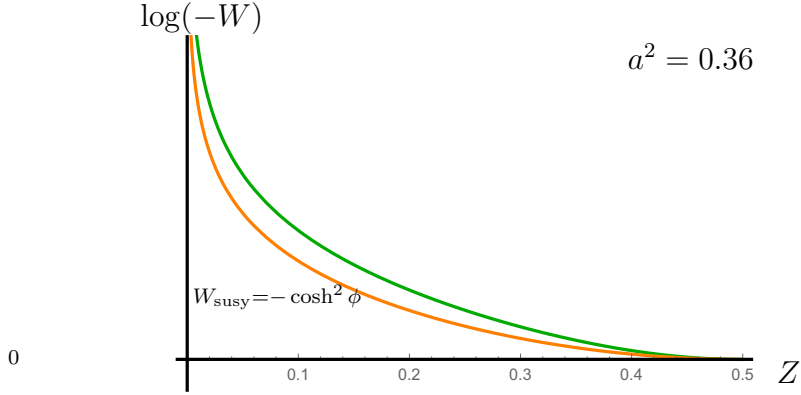
\begin{figure}[ht]
    \centering
    \includegraphics[height=4.8cm]{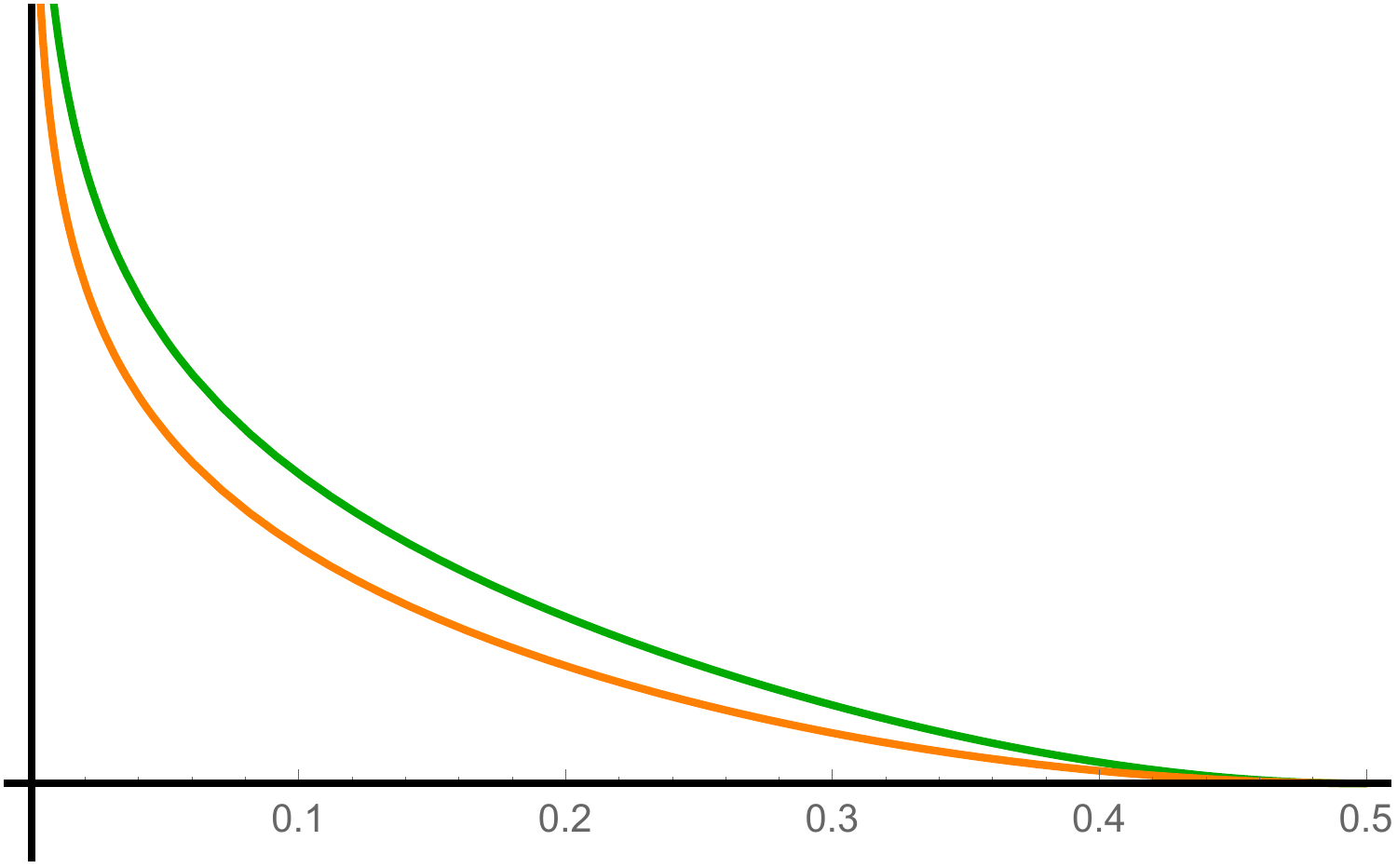}
    \begin{tikzpicture}[overlay]
        \node at (-7.5,5) (1) {$\log(- W)$};
        \node at (-6.5,1.3) (1) {$ \scriptstyle W_{\rm susy} = -\cosh^2 \f$};
        \node at (0.3,0.3) (1) {$Z$};
        \node at (-10,0.5) (1) {$\scriptstyle 0$};
        \node at (-1,4.5) (1) {$a^2 = 0.36$};
    \end{tikzpicture}
    \caption{(Logarithmic) plots of the superpotential $W_{\rm susy}= -\cosh^2 \f$ (orange) corresponding to the supersymmetric $W_-$-type flow from $p_1$ to $p_4$ found in \cite{Deger:2002hv}, and of the fake superpotential corresponding to the $W_+$-type flow from $p_1$ to $p_5$ started by $u_2$ (green). To make the figure more representative we plot $\log( -W)$ against $Z$.}
    \label{fig:WW0}
\end{figure}

Each of the vectors starts an RG flow in two directions depending on the sign of the constants $k_{1,2}$. Since $\D_+ > \D_-$ there exists a single flow where $\phi \sim e^{-\D_+ A}$, that is when $k_1=0$. This flow corresponds to a $W_+$-type solution, that is consistent with its uniqueness. On the other hand, there is a continuum of flows where $\f\sim e^{-\D_- A}$. These correspond to solutions of $W_-$-type. On Fig. \ref{fig:am12}  the  vectors $u_1$ and $u_2$  at the point $p_1$ are depicted by blue and black lines respectively. As it has been noticed in \cite{Golubtsova:2022hfk} and also can be seen on the Figure, the flow started by $u_1$ (highlighted by orange), that is a $W_-$-type solution, is precisely the supersymmetric flow of \cite{Deger:2002hv}. This is consistent with the previous Section. From Table \ref{tab:interpretations} we read that for Dirichlet boundary conditions (assumed in \cite{Deger:2002hv}) this flow is triggered by a single-trace deformation of the CFT at the critical point $p_1$. For Neumann boundary conditions the solution corresponds to RG flow of a (different) CFT at $p_1$ in a phase with a non-vanishing VEV of the scalar operator $\mc{O}_{\D_-}$.

The same is true for all flows inside the area bound by the red line on Fig. \ref{fig:am12} (whose origin will be clear in a moment) despite the flow highlighted by green, that is started by $u_2$ that is of $W_+$-type. As we see from the phase picture this flow ends at the critical point $p_5$, and its interpretation is inverse: for Dirichlet boundary conditions it is triggered by a non-zero VEV, for Neumann boundary conditions it is triggered by a single-trace deformation. To illustrate the difference between this flow and the $W_-$-type flow of \cite{Deger:2002hv} we plot $W$ for both solutions on Fig. \ref{fig:WW0}. For the latter the expression is analytical, while for the former we are only able to find a numerical solution to field equations. It is important to note here, that the difference is not merely an artifact of numerical calculation, since we check that the same calculation for the orange flow corresponding to the supersymmetric solution of \cite{Deger:2002hv} gives precisely the same curve as the analytical expression $W_{\rm susy} = - \cosh^2 \f$.

Lets us now comment on the area bound by the red line. Notice first, that each vector $u_1$ and $u_2$ can start flows in the directions opposite to those depicted on Fig. \ref{fig:am12}. However, as one learns from the phase picture all these flow to points $Z=1$, or equivalently $\phi \to -\infty$, that is the weak coupling limit on the gauge theory side. The weak gauge/gravity correspondence we are working in does not describe this regime since more information from the full quantum theory must be added on the gravity side. Hence, the allowed RG flows starting at $p_1$ are the ones inside the region bounded by the red line and the flow of \cite{Deger:2002hv}. Other solutions in general do not describe RG flows of a dual gauge theory. Formally, the flow along the red line itself that bounces the saddle points $p_7$ and $p_8$ belongs to the family of $W_-$-type, however, due to the same reasons one cannot think of it in terms of the dual theory. Hence, the closer a flow gets to the red line, the less control one has of the duality correspondence. For this reason, the points $p_{7,8,9}$ do not correspond to phase transition points of the theory whose RG flow starts at $p_1$, rather they illustrate the bounds imposed by physics of the correspondence.

Note also the rather counter intuitive position of the $W_+$-type solution inside the red-line area. Indeed, as it has been shown in \cite{Papadimitriou:2007sj} in general $W_+$-type solutions are expected to be a limit of the family of $W_-$-type solutions. Hence, simply from the Fig. \ref{fig:am12} one would intuitively expect that the orange flow (along $u_1$) must be of the $W_+$-type, since it bounds the flows inside the red-line area. However, the analysis shows that this is not the case and the boundary flow is of the $W_-$-type. Furthermore, explicit expression of the corresponding solution is known \cite{Deger:2002hv} and it fits to the above picture.

Let us also comment on the $c$-theorem, stating that central charge in a CFT must decrease when going to the IR. On the gravity side for the domain wall backgrounds the $c$-function determining the central charges is defined as \cite{Henningson:1998gx,Freedman:1999gp}
\begin{equation}
    c = \fr{C_0}{\dot{A}},
\end{equation}
where $C_0$ is some negative constant. Now, for its derivative we have
\begin{equation}
    \fr{dc}{dA} = \fr{\dot c}{\dot A} = \fr{C_0}{\dot A} \fr{X^2}{a^2 }.
\end{equation}
Since $\dot A = - W$ is always positive \eqref{wexpansion}, we see that the $c$-function is decreasing along the flow as long as $X$ is nonzero. Yet, there exist flows crossing the points where $X^2=0$ for which the first derivative of the $c$-function vanishes. However, direct check shows that the second derivative is $d^2c/dA^2 \sim X$ and hence vanishes at these points as well, meaning that $X^2=0$ are inflection points of the $c$-function. We conclude that the $c$-function is a monotonic function and the $c$-theorem holds for the RG-flows considered in this section. We will comment more on the flows crossing the line $X^2=0$ in Section \ref{sec:exotic} which is devoted to bouncing RG-flows.

\subsubsection{\texorpdfstring{$1/2<a^2 <1$}{1/2<a2<1}}

Let us now turn to the range of the parameter $a^2$ when the potential $V(\f)$ develops additional (AdS) minima at finite values of $\phi$. The corresponding solutions have been analyzed in \cite{Deger:2002hv} for Dirichlet boundary conditions. Before proceeding with analysis of the full picture of flows let us mention some troubles with the flow equations written in the form \eqref{eqZX}. Notice that the term proportional to $V^{-1}$ in the second line of \eqref{eqZX} develops singularity at points when $V=0$. For $0<a^2 \leq 1/2$ the potential \eqref{eq:potential} does not have zeroes and hence the phase picture is free from singularities and all flows in terms of $(X,Z)$ are well defined and can be put in one-to-one correspondence to solutions of the initial field equations. This however is no longer the case for $1/2 < a^2 < 1$ when the zeroes are given by
\begin{equation}
    V(\f_0) = 0 \,, \,  \cosh^2 \f_0 = \fr{2a^2}{2a^2-1}.
\end{equation}

\begin{figure}[ht]
    \centering
    \includegraphics[height=8cm]{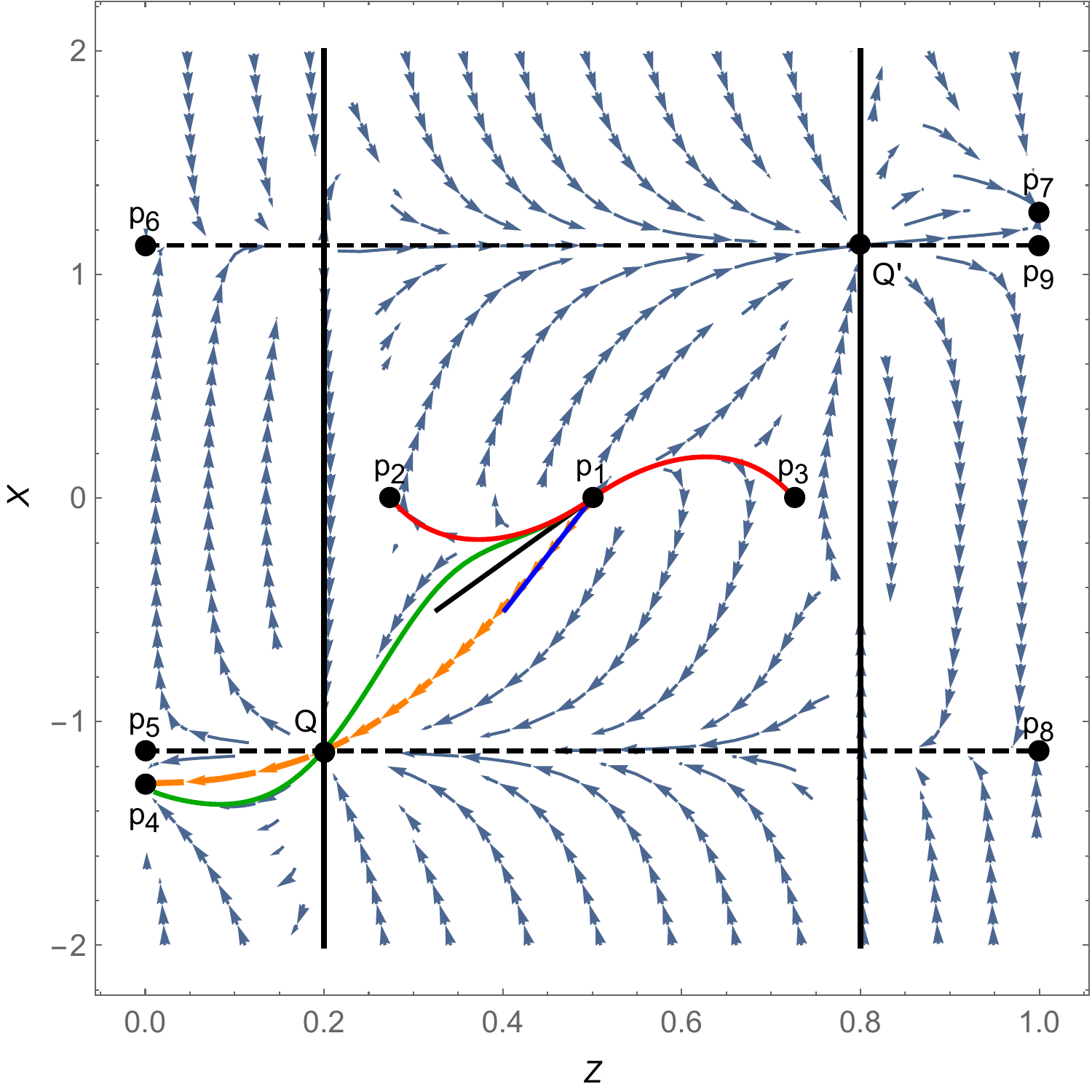}
    \caption{Phase flow for $1/2<a^2 <1$ ($a^2 = 0.64$). Vertical solid lines denote points where $V=0$ and hence phase flows are singular. Horizontal dashed lines denote points $X^2 = 2 a^2$, four points of their intersection with the solid lines are regular. Two flows studied in \cite{Deger:2002hv} are depicted by orange and red lines. The red one corresponds to non-supersymmetric flows interpolating between two different AdS vacua (from $p_1$ to either $p_2$ or $p_3$) whereas the orange curve is supersymmetric.}
    \label{fig:ag12}
\end{figure}

It is important to note that these are not the critical points of the potential and hence $V'(\f_0)\neq 0$, hence the vector flow  naively develops a singularity. Now, from the relation
\begin{equation}
    \label{eq:dotA2V0}
    4 V = \dot A^2 \left(\fr{X^2}{a^2} - 2\right).
\end{equation}
we see that at $V=0$ either $\dot A=0$ or $X_0^2 = 2a^2$ (or both). The second option is safe since the zeroes in the flow equation cancel rendering regular phase flow on the lines $X_0 = \pm \sqrt{2}a$. On Fig. \ref{fig:ag12} these are depicted by dashed horizontal lines $p_6-p_9$ and $p_5-p_8$. Since the potential $V$ is a symmetric function of $\phi$ there are two lines $Z=Z_1$ and $Z=Z_2$ where $V=0$, these are depicted by solid vertical lines on Fig. \ref{fig:ag12}.

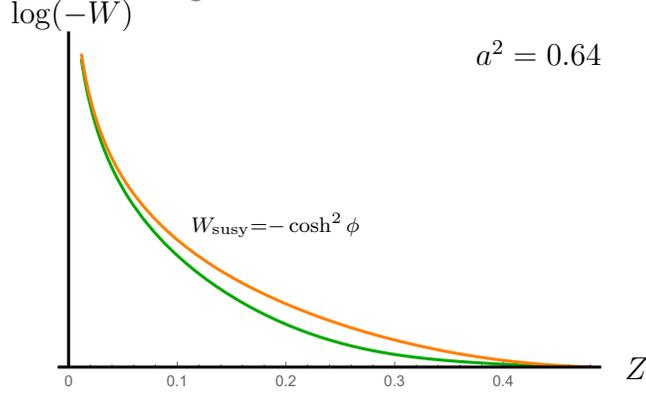
\begin{figure}[ht]
    \centering
    \includegraphics[height=4.8cm]{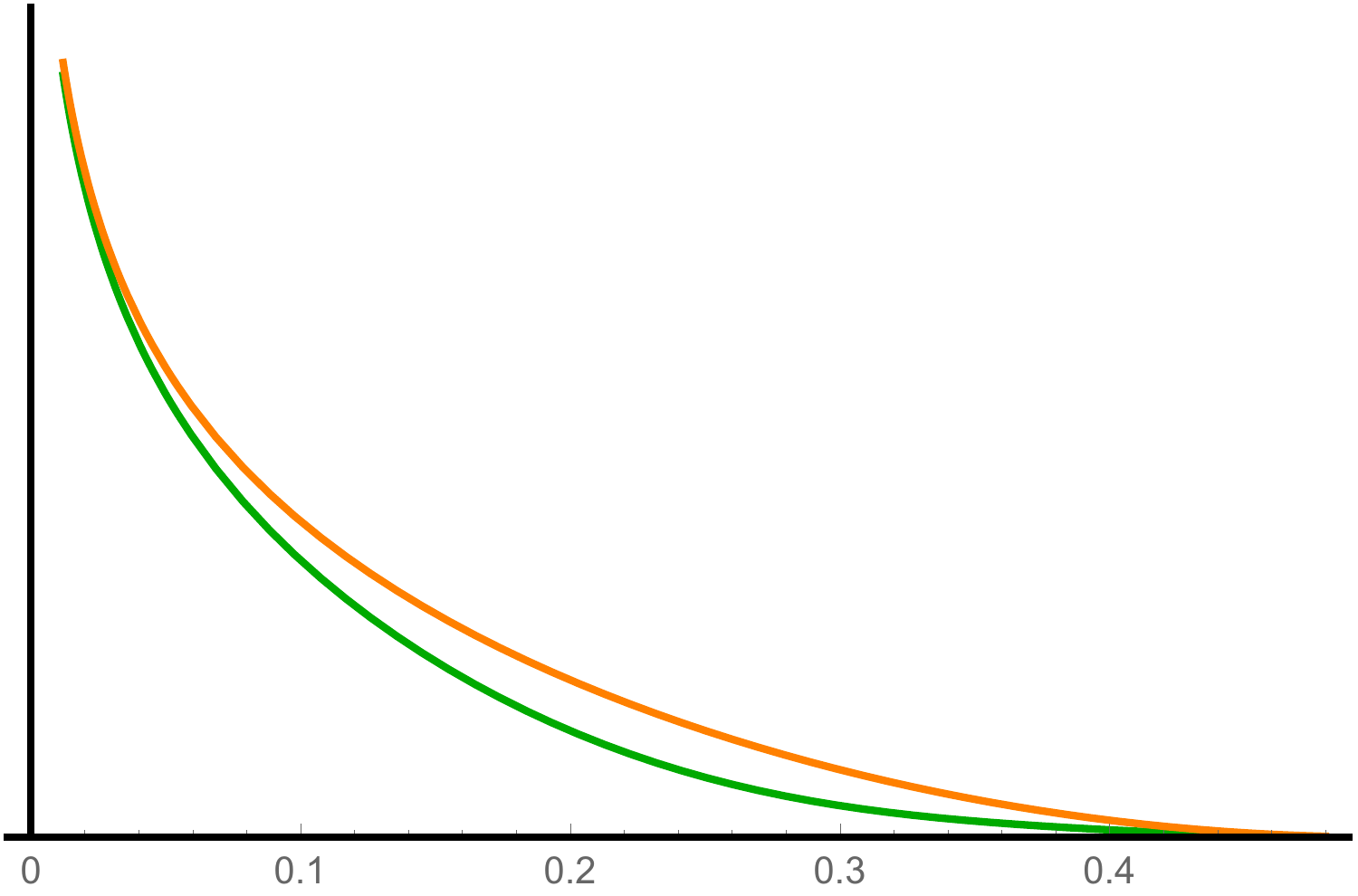}
    \begin{tikzpicture}[overlay]
        \node at (-7.2,5) (1) {$\log(- W)$};
        \node at (-4.5,2.2) (1) {$ \scriptstyle W_{\rm susy} = -\cosh^2 \f$};
        \node at (0.3,0.3) (1) {$Z$};
        \node at (-1,4.5) (1) {$a^2 = 0.64$};
    \end{tikzpicture}
    \caption{(Logarithmic) plots of the superpotential $W_{\rm susy}= - \cosh^2 \f$ (orange) corresponding to the supersymmetric $W_+$-type flow found in \cite{Deger:2002hv}, and of the fake superpotential corresponding to the $W_-$-type flow (green) starting again from $p_1$ and flowing to the same fixed point $p_4$ passing the same regular point $Q$. To make the figure more representative we plot $\log( -W)$ against $Z$.}
    \label{fig:WW0g12}
\end{figure}

For the potential  \eqref{eq:potential} there are four points $(Z_{1,2},\pm X_0)$ on the singularity lines where phase flows are regular. As we learn from the phase picture on Fig. \ref{fig:ag12} there are three types of flows starting from $p_1$. The first type is represented by the flows starting at $p_1$ and ending at either $p_2$ or $p_3$ (drawn by red lines on the figure), that correspond to solutions interpolating between two different AdS vacua. These were analyzed in \cite{Deger:2002hv}. The second and third types are given by flows going to the regular point $Q$ and $Q'$ respectively. As we have discussed above we do not consider the latter since the weak gauge/gravity correspondence breaks when $Z$ tends to unity. 

Let us consider flows of the second type in more details. These start at $p_1$ and all intersect at the regular point $Q$. In general crossing of phase trajectories of an autonomous system can happen at unstable fixed point, good example of which is the upper position of a pendulum. Hence, one thinks of $Q$ as of an additional unstable  fixed point developed by the system, after entering which the system can further develop along any outgoing trajectory. However, there exists an exact solution to field equations, which goes from $p_1$ to $p_4$ for any choice of $a^2$ (orange flow on \ref{fig:ag12}) successfully passing the unstable point $Q$. Interestingly enough, numerical solution of the autonomous system equation for $Z$ and $X$ allow to find many other examples of flows starting at $p_1$ and ending at $p_4$ passing the regular point $Q$, one of which is depicted on Fig. \ref{fig:ag12} by the green line. As before these correspond to two different functions $W = W(Z)$, as it can be seen from Fig. \ref{fig:WW0g12}. Since the flows start and end at the same points, the fake superpotential corresponding to the second flow (green) coincides with the true superpotential at the beginning and at the end of the flow. We conclude that for this flow (at least part) of supersymmetry is restored at the IR.

Notice that due to the relation \eqref{eq:dotA2V0} everywhere on the vertical walls except four intersection points the function $\dot A$ vanishes. It means that at the singularity lines the function $A=A(r)$ is not a monotonic function of $r$ and hence cannot be used as a scale factor. Hence, these lines are not simply the result of a bad choice of functions $X$ and $Z$ and are rather the intrinsic property of the model, which includes having $A=A(r)$ as the scale factor. On principle one might expect that introducing back the fields of the model that we truncated, namely the scalar field $\chi$ (phase of the complex scalar field) and the vector field  $A_\mu$ the full phase picture becomes regular. We leave this for further investigation.

We conclude this subsection by analysis of behavior of the flows near the critical point $p_1$. For $2a^2 > 1$ a general solution starting from $p_1$ has the form
\begin{equation}
    \begin{bmatrix}
        Z\\
        X
    \end{bmatrix}
    =
    \begin{bmatrix}
        1/2\\
        0
    \end{bmatrix}
    + k_1 e^{- \D_+ A} u_1 + k_2 e^{-\D_- A} u_2.
\end{equation}
Since $\D_+ > \D_-$ and $A \to \infty $ when approaching the critical point there is only one trajectory where $e^{-\D_+ A}$ dominates, that is $k_2 = 0$. On the other hand, there exist an infinite family of trajectories behaving as $e^{-\D_- A}$ near the critical point. 

The $W_+$-type flow starting at the point $p_1$ and ending at $p_4$ is the solution of \cite{Deger:2002hv}. All other flows starting from $p_1$ (including the ones ending at $p_{2,3}$) are of the $W_-$-type. Hence, the flows interpolating between two AdS vacua (two different CFT's) are triggered by  a single-trace deformation in the Dirichlet case and by a phase with non-vanishing VEV in the Neumann case. These are also compatible with mixed boundary conditions, in which case the corresponding CFT is a multitrace deformation of the CFT with Neumann boundary conditions.

\subsubsection{\texorpdfstring{$1 \leq a^2$}{1 < a2}}

When $1 \leq a^2$ the AdS critical points $p_{2,3}$ disappear, while the singularities of $1/V(\f)$ have the same structure as discussed above. The general picture of the RG flows passing the regular point $Q$ is the same as in the case $1/2 < a^2 < 1$.

\subsubsection{\texorpdfstring{$a^2=1/2$}{a2=1/2}}

Finally, let us consider the special value $a^2 = 1/2$, for which there is only one critical point $\phi_{*1}$ and the
conformal dimensions are equal $\D_+ = \D_- = 1$ \eqref{dimension}. Moreover, the scalar field near the AdS boundary saturates the BF bound \eqref{BF}. In the language of the autonomous system this corresponds to merging of eigenvalues and hence, there is a single eigenvector $u_1$, that is the one corresponding to the eigenvalue $\lambda = -1$:
\begin{equation}
    \mc{M} u_1 = - u_1.
\end{equation}
The second vector defining evolution of the system is given by the adjoint eigenvector $u_2$ satisfying
\begin{equation}
    (\mc{M} - \lambda \mathbf{1})\,u_2 = u_1,
\end{equation}
where $\lambda=-1$ is the eigenvalue of $u_1$ and $\mathbf{1}$ denote the unity matrix. For the critical point $p_1$ we have explicitly
\begin{equation}
    \begin{aligned}
        &u_1 = 
            \begin{bmatrix}
                1\\
                4
            \end{bmatrix}, &&
        u_2 = 
            \begin{bmatrix}
                \a\\
                4 \a - 1
            \end{bmatrix},
    \end{aligned}
\end{equation}
where $\a$ is an arbitrary number. For numerical calculations we choose it to be $\a = 1/4 $.

\begin{figure}[H]
    \centering
    \includegraphics[height=8cm]{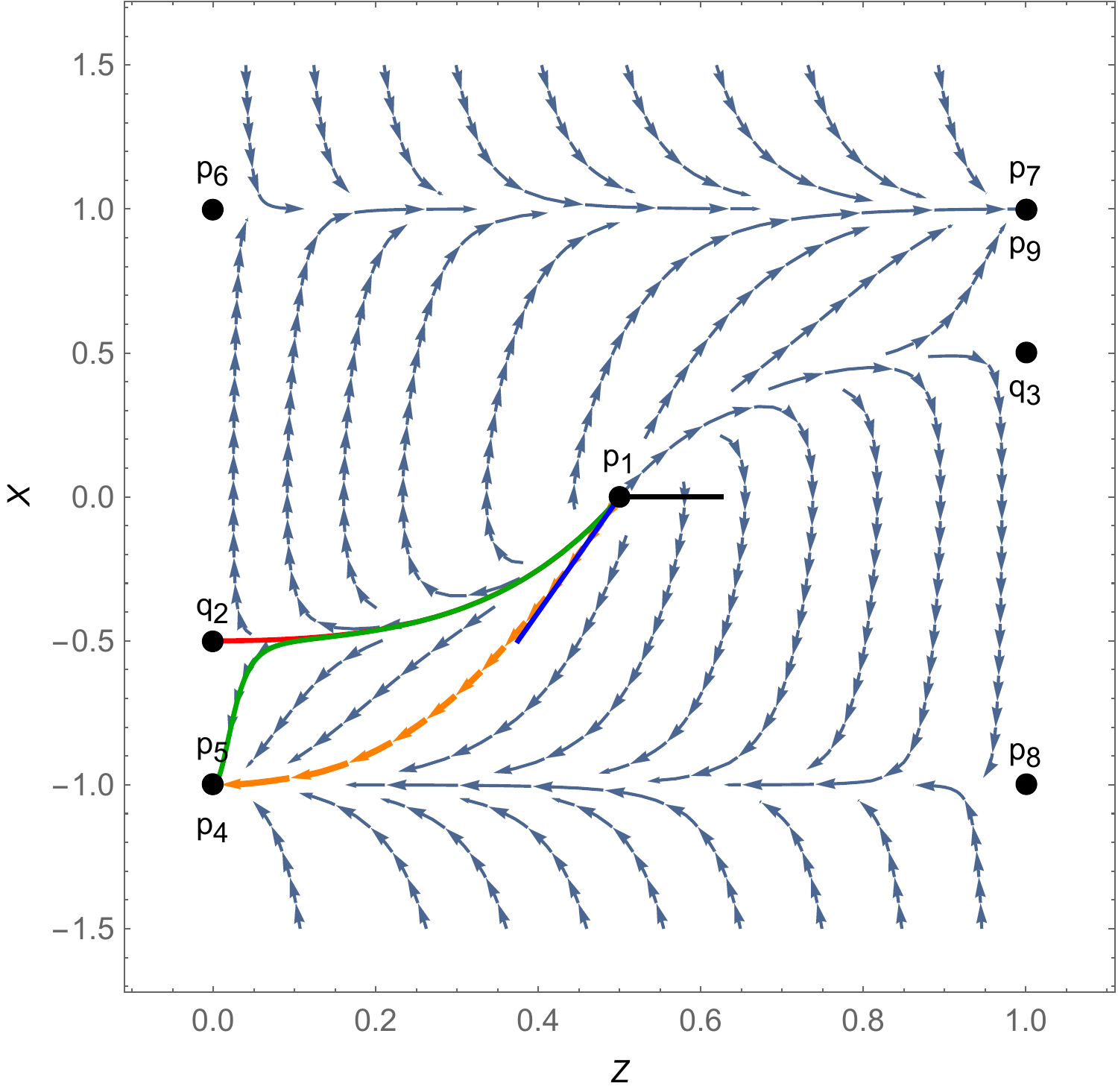}
    \caption{Phase flow for $2a^2=1$. The flow given by the exact solution \eqref{scafieldDeg},\eqref{scafDeg} of \cite{Deger:2002hv} is depicted by orange arrows. Two additional flows are given by the green and the red lines. Both start at $p_1$, the former ends at the stable node $p_4$, the latter ends at the new unstable node $q_2$.}
    \label{fig:ae12}
\end{figure}

A general solution to the flow equations near the critical point is then given by
\begin{equation}
\label{eq:ZXfor05}
    \begin{bmatrix}
        Z\\
        X
    \end{bmatrix}
    =
    \begin{bmatrix}
        1/2\\
        0
    \end{bmatrix}
    + (k_1+k_2A) e^{- \D A} u_1 + k_2 e^{-\D A} u_2,
\end{equation}
where as before $k_1$ and $k_2$ are arbitrary constants defining the type of the solutions.

For the critical points of the system we have the following. First, the points $p_4$ and $p_5$ merge as well as $p_7$ and $p_9$. Second, the point $p_2$ and $p_3$ that correspond to AdS vacua for $1/2 < a^2 <1$ do not exist for $a^2=1/2$. Instead, the system develops new critical points $q_{2} = (0,-1/2)$ and $q_3 = (1,1/2)$ that are saddle points. Note, that the point $p_1$ is now a degenerate unstable node. The points $p_{4,5}$ and $p_{7,9}$ are still stable nodes and attract phase curves.

\begin{figure}[ht]
    \centering
    \includegraphics[height=4.8cm]{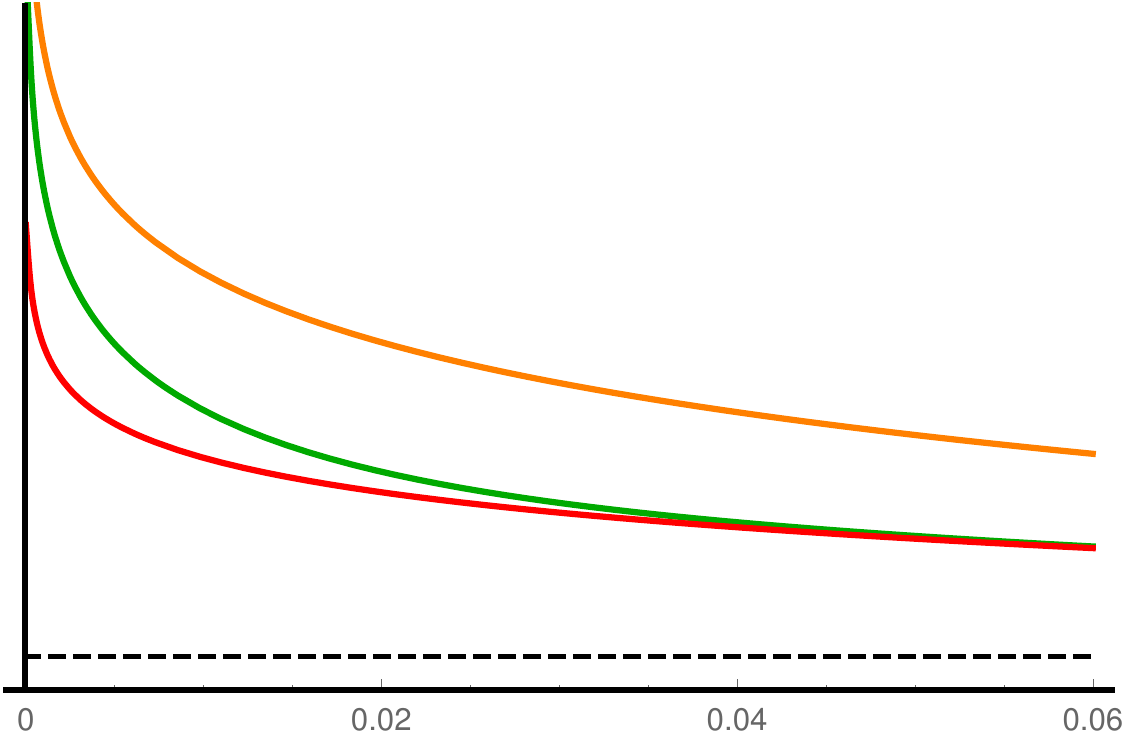}
    \begin{tikzpicture}[overlay]
        \node at (-7.2,5) (1) {$\log(- W)$};
        \node at (-5.5,3.5) (1) {$ \scriptstyle W_{\rm susy} = - \cosh^2 \f$};
        \node at (0.3,0.3) (1) {$Z$};
        \node at (-6,0.7) (1) {$\scriptstyle W=-1$};
        \node at (-1,4.3) (1) {$a^2 = 0.5$};
    \end{tikzpicture}
    \caption{(Logarithmic) plots of the superpotential $W_{\rm susy}= - \cosh^2 \f$ (orange) corresponding to the analytical $W_+$-type flow found in \cite{Deger:2002hv}, and of the fake superpotential corresponding to the $W_-$-type flow (green) from $p_1$ to the same fixed point $p_4$. The dashed line corresponds to $W=-1$. Red line denotes the $W_-$-type flow from $p_1$ to $q_2$. To make the figure more representative we plot $\log( -W)$ against $Z$.}
    \label{fig:WW0e12}
\end{figure}
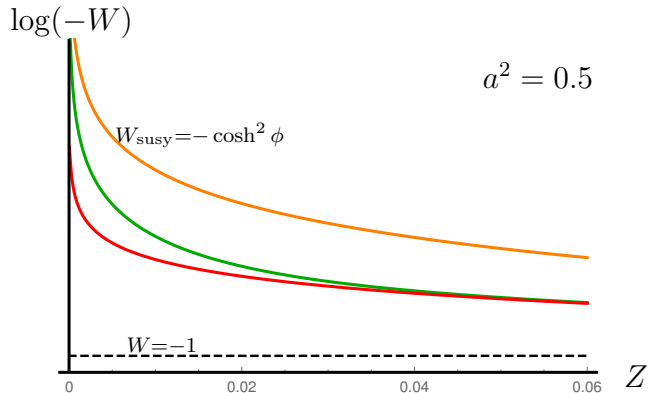

From the general behavior \eqref{eq:ZXfor05} of the scalar field and its derivative we see that solutions with $k_2 \neq 0$ dominate those with $k_2=0$. We will refer to the former as the log-type solutions since they depend linearly on $A \sim \log z$, where $z$ is the standard AdS radial coordinate. As we learn from the phase picture on Fig. \ref{fig:ae12} the exact solution of \cite{Deger:2002hv} corresponds to $k_2 = 0$ and hence is a distinguished solution. For Dirichlet boundary conditions this solution corresponds to RG flow triggered by a non-vanishing VEV. Indeed, in this case for the scalar field we have the expansion
\begin{equation}
    \f(r,x) = e^{-r}\Big[-2 r (\f_{(0)}(x) + \dots ) + \tilde{\f}_{(0)}(x) + \dots\Big],
\end{equation}
where the field $\f_{(0)}(x)$ is the source on the boundary and $\tilde{\f}_{(0)}(x)$ gives the corresponding expectation value. For the solution in question we do not have the linear term while the other term is non-vanishing. For Neumann boundary conditions the roles of the fields are interchanges and hence the RG flow is triggered by a single-trace deformation. Hence, one may use the $W_+$ column of Table \ref{tab:interpretations}.

Similarly for all other flows one uses the $W_-$ column of Table \ref{tab:interpretations}. In particular we explicitly draw two distinct flows. One (green) starts from $p_1$ and ends at the stable point $p_4$, the other (red) starts at $p_1$ and ends at the new unstable critical point $q_2$. Both of them are of the log-type and the corresponding fake superpotentials are presented at Fig. \ref{fig:WW0e12}. We draw the plots only near the region $Z=0$ to illustrate that the fake superpotential for the green flow ($p_1-p_4$) asymptotically approximates the true superpotential, that is expected since it ends up at the same fixed point. To the contrary the fake superpotential of the red flow behaves differently since the solution tends to a different critical point. Certainly near $Z=0.5$ all three get asymptotically close to the line $W=-2$, however green and red flows become asymptotically close much earlier. This can also be seen from the phase picture.

\subsection{Exotic flows}
\label{sec:exotic}

The holographic RG prescription implies that every solution (satisfying the $c$-theorem) describes an RG flow between UV and IR critical points. In general from the standard perturbative QFT point of view one would expect this flow to be monotonic, i.e. VEV of an operator monotonically changes from its value in the UV to that in the IR. However, from the phase pictures presented we find that this is not completely the case for the system in question as there exist a set of flows for which $Z=Z(A)$ is not a monotonic function. Certainly, this is not a complete surprise and a classification of holographic RG flows has been already considered in \cite{Kiritsis:2016kog}. According to the classification, in addition to the so-called limit cycles one may have bouncing flows, when the RG evolution turns back at some value of VEV, and skipping solutions, when the flow skips one or several critical points. Although there has not been much progress in construction of such exotic RG flows from the conventional QFT point of view, in \cite{Curtright:2011qg} it has been shown that for example limit cycles do not necessarily break $c$-($a$-)theorem. The reason is that although the superpotential bounces as a function of VEV, it is a monotonic function of the scale.

\begin{figure}[ht]
    \centering
    \includegraphics[height=8cm]{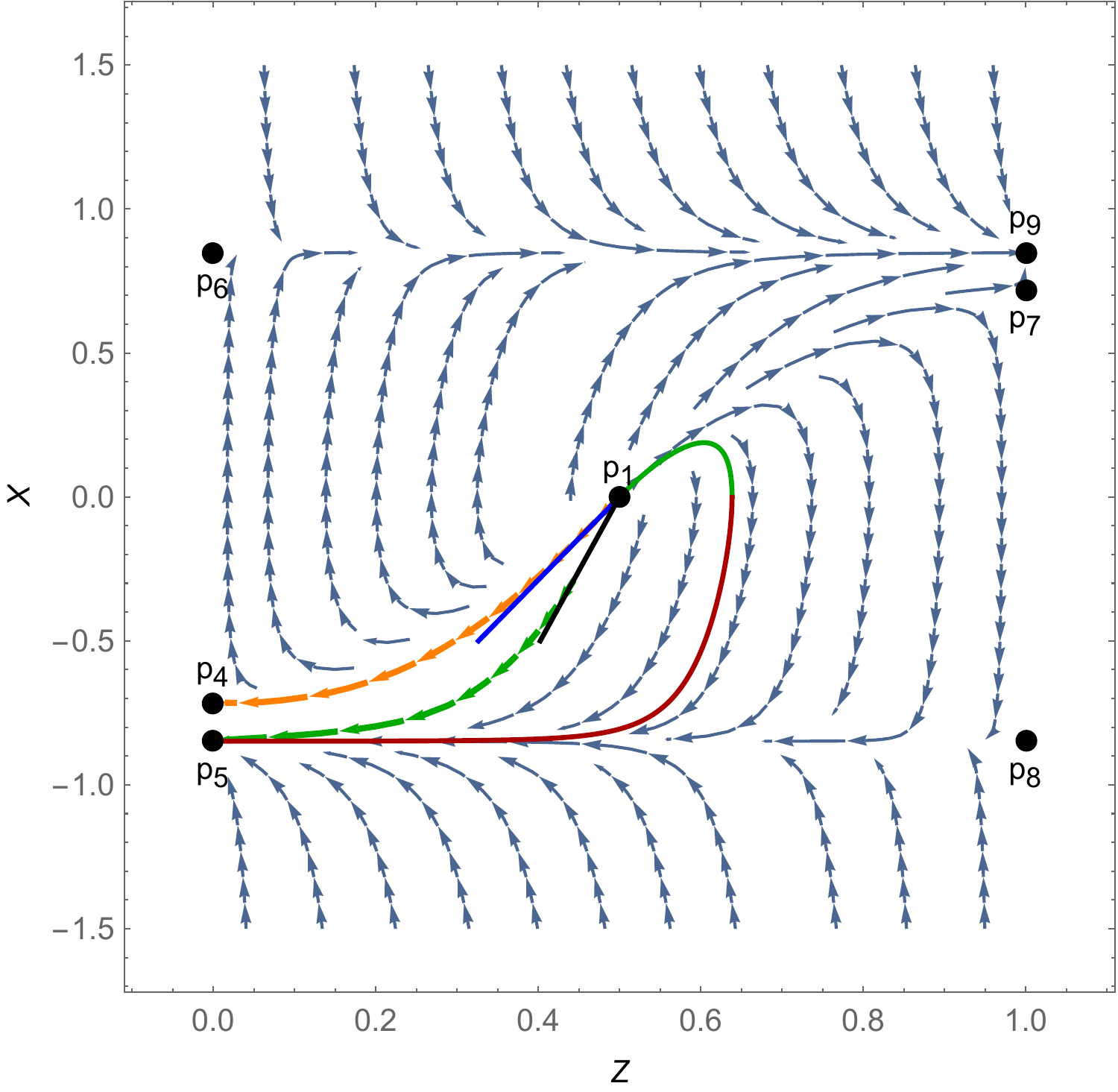}
    \caption{Phase flow for $a^2=0.36$. The flow given by the exact solution of \cite{Deger:2002hv} is depicted by orange arrows. A monotonic flow between the critical points $p_1$ and $p_5$ is depicted by green arrows. A bouncing flow between $p_1$ and $p_5$ is depicted by the solid line of two colors, green and red.}
    \label{fig:exotic}
\end{figure}

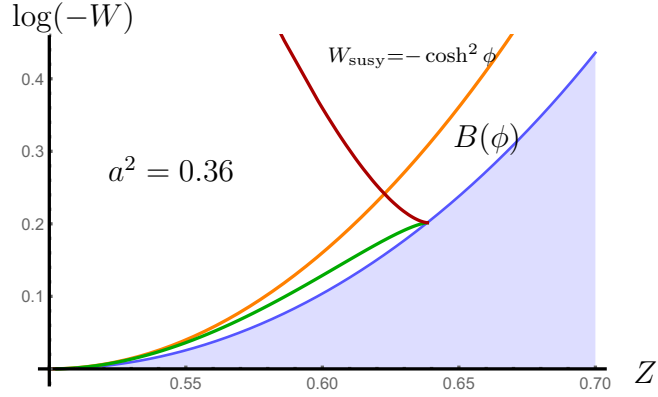
\begin{figure}[ht]
    \centering
    \includegraphics[height=4.8cm]{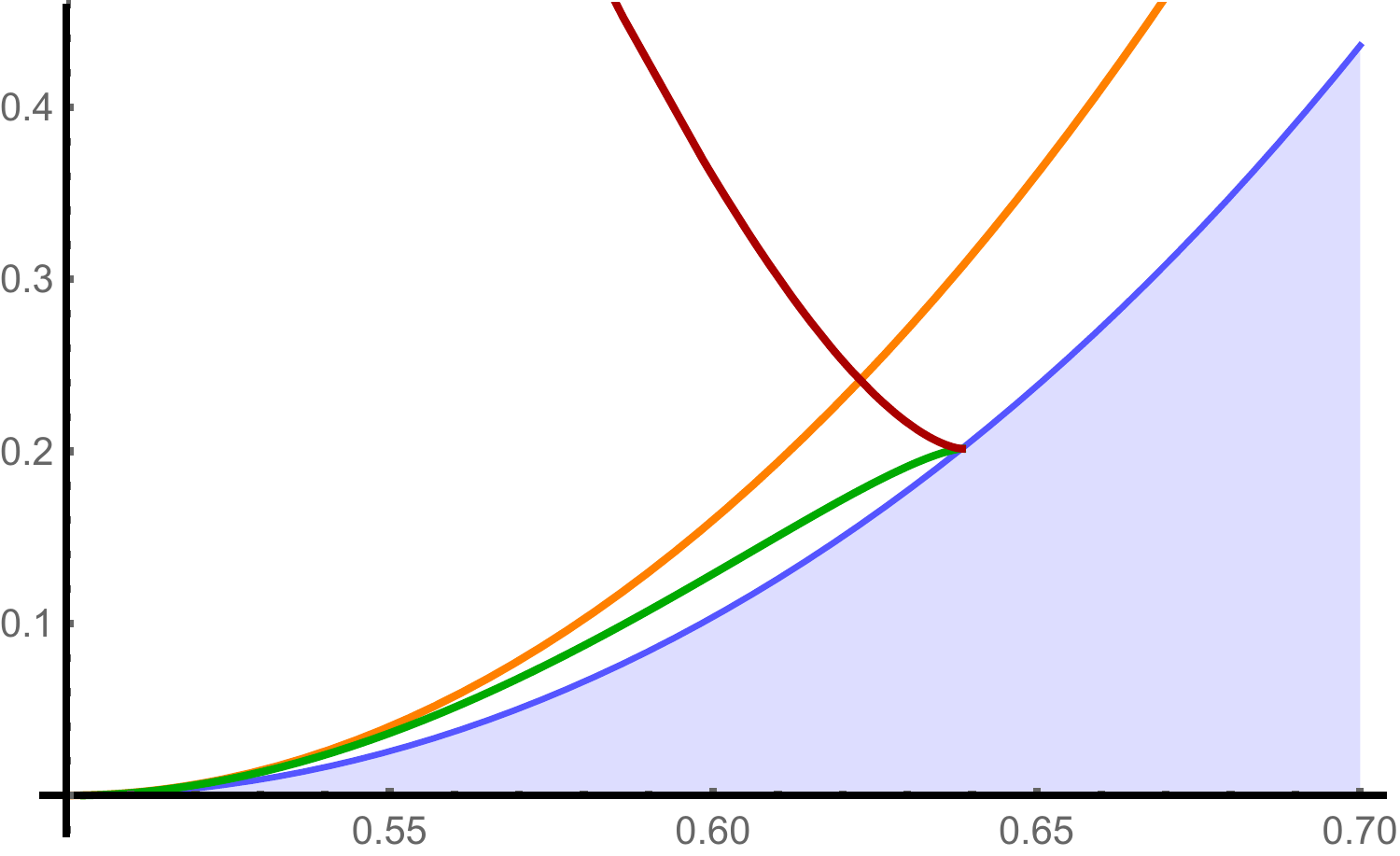}
    \begin{tikzpicture}[overlay]
        \node at (-7.3,5) (1) {$\log(- W)$};
        \node at (-2.8,4.5) (1) {$ \scriptstyle W_{\rm susy} = - \cosh^2 \f$};
        \node at (0.3,0.3) (1) {$Z$};
        \node at (-1.8,3.4) (1) {$B(\phi)$};
        \node at (-6,3) (1) {$a^2 = 0.36$};
    \end{tikzpicture}
    \caption{(Logarithmic) plots of the superpotential $W_{\rm susy}= - \cosh^2 \f$ (orange) corresponding to the supersymmetric $W_+$-type flow found in \cite{Deger:2002hv}, and of the fake superpotential corresponding to the bouncing solution depicted on Fig. \ref{fig:exotic}. The red branch of the bouncing superpotential asymptotically converges with the curve $B(\phi)$ at $Z=0$.}
    \label{fig:Wexotic}
\end{figure}

Let us focus on the bouncing RG flows, that we observe in our case, and for concreteness let $2a^2 < 1$. From the gravitational point of view the reason for bouncing solutions to exist is that the flow equation is first order while the field equations are second order, and hence $W$ may have  extrema that do not coincide with that of $V$. Therefore, the equation \eqref{VthroughW} has in principle two branches of solutions: a growing solution $W_{\uparrow}(\phi)$ and a decreasing solution $W_{\downarrow}(\phi)$ (these should not be confused with $W_\pm(\phi)$, that are solutions with different asymptotics in the UV). A bounce happens when two solutions intersect. Indeed, from \eqref{VthroughW} one finds that $W$ is bounded from above by a curve $B(\phi)$
\begin{equation}
    B(\phi) = \sqrt{-2 V(\f)}.
\end{equation}
On the other hand, one may rewrite the equation \eqref{VthroughW} as follows
\begin{equation}
    W'(\f) = \pm \fr{\sqrt{2}}{a}\sqrt{W^2 - B^2}.
\end{equation}
This implies that at the extrema points, $W'(\phi_*)=0$, the superpotential $W$ hits the curve $B$. Now, if this is not the end of the corresponding RG flow, i.e. $V'(\phi_*)\neq 0$, evolution must proceed further which will make $W$ complex. To stay in the set of real values of the superpotential we must glue the decreasing solution to the corresponding increasing solution, that has an extremum at the same point.

\begin{figure}[ht]
    \centering
    \includegraphics[height=4.8cm]{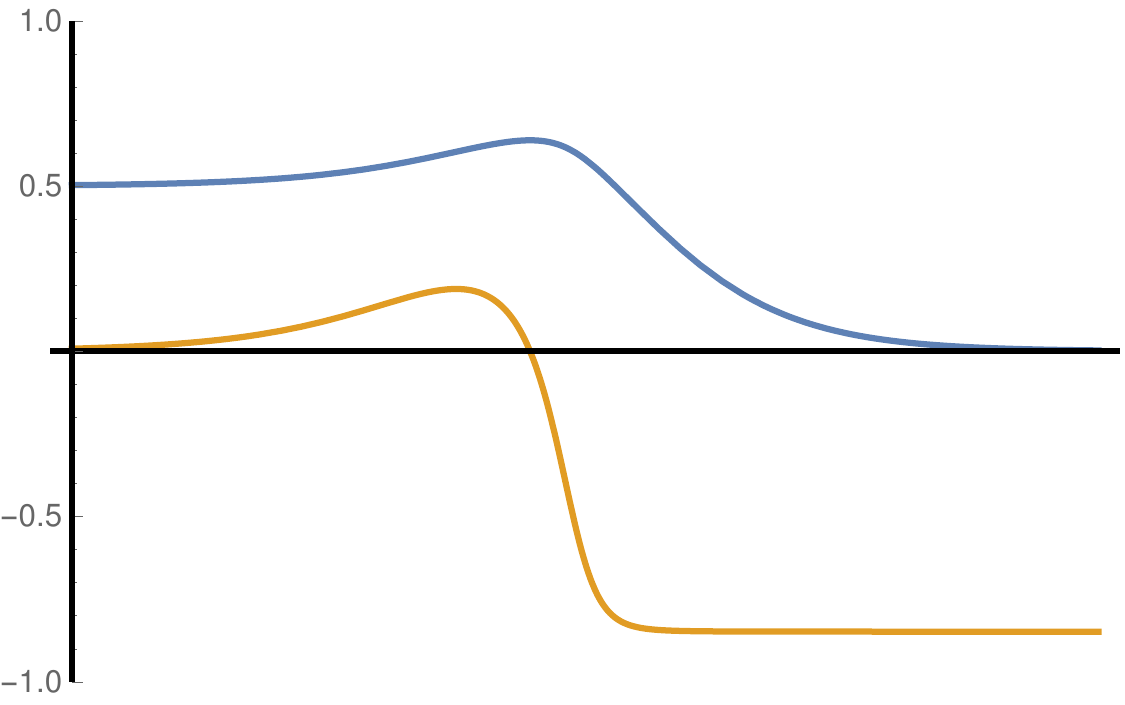}
    \begin{tikzpicture}[overlay]
        \node at (-6,3.8) (1) {$Z(A)$};
        \node at (-6,2.8) (1) {$ X(A)$};
        \node at (0.1,2.5) (1) {$A$};
        \node at (-1,3.5) (1) {$a^2 = 0.36$};
    \end{tikzpicture}
    \caption{Plots of the functions $X(A)$ and $Z(A)$ for the bouncing solution. RG evolution goes from the left to the right.}
    \label{fig:XZexotic}
\end{figure}
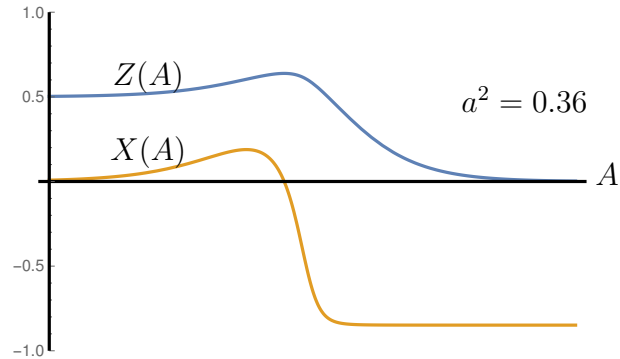

On the Fig. \ref{fig:exotic} a solution corresponding to bouncing RG flow is depicted by the solid line of two colors. We obtain this phase flow by numerically solving the flow equations, parametric plot of the obtained solution $(Z(A),X(A))$ gives the desired flow from $p_1$ from $p_5$. As it is shown on Fig. \ref{fig:XZexotic} both functions have a maximum at certain (different) scales, hence, both inverse functions $A(X)$ and $A(Z)$ have two branches as the (beta-)function $X=X(Z)$. These two branches are depicted by different colors on Fig. \ref{fig:exotic}. 

\begin{figure}[ht]
    \centering
    \includegraphics[height=4.86cm]{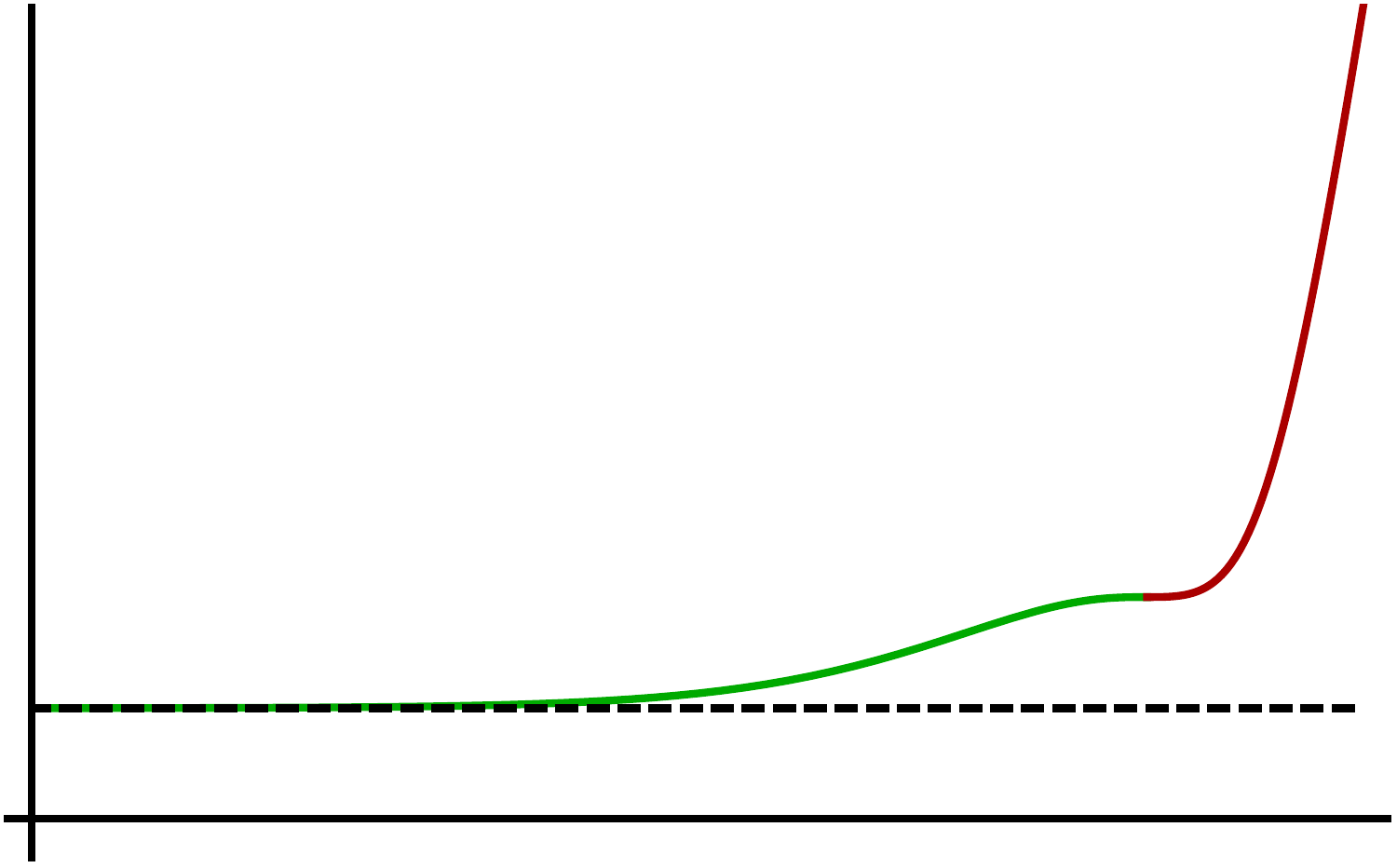}
    \begin{tikzpicture}[overlay]
        \node at (-7.5,5.1) (1) {$\log(-W)$};
        \node at (-0.3,1.2) (1) {$ \scriptstyle W=-1$};
        \node at (0.2,0.3) (1) {$A$};
        \node at (-2,4.5) (1) {$a^2 = 0.36$};
    \end{tikzpicture}
    \caption{Plot of the fake superpotential for the bouncing solution against the scale factor (RG evolution goes from the left to the right).}
    \label{fig:WAexotic}
\end{figure}
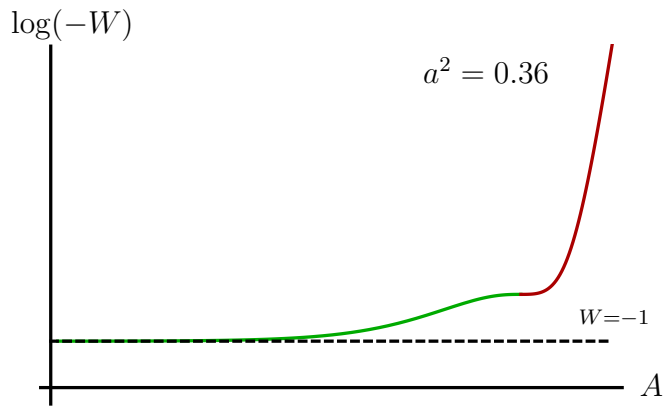

Certainly, although the function $W=W(Z)$ is defined via two branches, due to the behavior of the function $Z=Z(A)$ the function $W=W(A)$ is monotonic, as required by the $c$-theorem. To check this we plot on Fig. \ref{fig:WAexotic} numerically $W=W(Z(A))$ using the interpolation functions we have found before.

We conclude that, the bouncing solution that we have analyzed here indeed respects the $c$-theorem, as expected. Moreover, looking at the phase curve one concludes that RG flows corresponding to such solutions are rather conventional for the system in question. Certainly, the very existence of such solutions might be a consequence of the truncation $A_\m=0$, $\Phi = |\Phi|$ we made from the very beginning. In other words, it could be that such form of bouncing solutions is simply the result of a projection of the full phase picture to the slice corresponding to the truncation. We discuss more on that in conclusions.

\section{Discussion}
\label{sec:disc}

In this work renormalization group flows of CFT's holographically dual to Poincar\'e domain wall solutions of the $D=3$ $\mc{N}=(2,0)$ gauged supergravity of \cite{Deger:1999st,Deger:2002hv} with various boundary conditions have been studied. For simplicity we focused on the case where the scalar coset space is $\rmSU(1,1)/\rmU(1) = \mathbb{H}^2$ and considered a truncated  subsector of the theory where vector field and phase of the scalar field vanish. To analyze RG flows we followed the approach in which supergravity field equations are represented as a dynamical system equations for the remaining real scalar field and its derivative w.r.t. the scale factor $A$ (VEV and beta-function respectively). We then constructed phase pictures for all allowed values of the parameter $a^2$ which is related to the curvature of the scalar coset space. The near boundary behavior of a solution allows to determine its type, $W_+$ or $W_-$ according to expansion of the corresponding fake superpotential. Adding to this the information on boundary conditions we interpret the obtained RG flows following the holographic dictionary of \cite{Papadimitriou:2004rz,Papadimitriou:2007sj}. For the most indicative representatives of RG flows we numerically solved dynamical equations and plotted the corresponding fake superpotentials. Among these are the supersymmetric and non-supersymmetric flows found in \cite{Deger:2002hv}. The results presented here complete the analysis of \cite{Golubtsova:2022hfk}. 

There are several aspects of these RG flows that are worth mentioning. First of all, from our phase pictures of the RG flows we conclude that for any allowed value of $a^2$ the system develops the so-called exotic flows, that in our case correspond to bouncing solutions. Here bouncing means that the corresponding fake superpotential as a function of the scalar field reflects from the curve that demarcates its real values from  complex (see Fig. \ref{fig:Wexotic}). Such bouncing flows fit the classification of \cite{Kiritsis:2016kog} and do not violate the $c$-theorem since the bouncing fake superpotential is a monotonic function of the scale parameter $A$ (see Fig. \ref{fig:WAexotic}). On the QFT side such bouncing exotic flows correspond to beta-functions represented by two branches and to flows that reverse their direction. To our knowledge such flows have not been constructed from actual QFT computations, however a similar behavior represented by limit RG cycles have been intensively studied for some time starting from the original paper \cite{Wilson:1974mb}. Then, in \cite{Gukov:2016tnp} the relation with dynamical systems was given which was followed by the observation of such cycles in quantum mechanical systems in \cite{Glazek:2002hq,Braaten:2004pg,Gorsky:2013yba,Dawid:2017ahd} (for a review see \cite{Bulycheva:2014twa}). Certain progress on RG cycles in the context of perturbative QFT has been reported recently in \cite{Jepsen:2020czw}. Although similar, our bouncing RG flows may be just an artifact of the truncation to the single scalar sector that we used. To understand this better, one has to construct a dynamical system corresponding to the complete set of supergravity field equations of our model \eqref{fullmodel} keeping also the phase of the scalar field but to do that one also has to turn on the gauge field which makes the analysis technically difficult. A more suitable model to understand this question is the $D=3$, $\mc{N}= 4$ gauged supergravity studied in \cite{Deger:2019jtl}. This model is a truncation of the supergravity that comes from $D=6$, $\mc{N}= 1$ supergravity via a consistent 3-sphere reduction \cite{Deger:2014ofa}. It contains 3 scalar fields but it is allowed to activate only two of them without any gauge fields. Two supersymmetric domain wall solutions for this case were also obtained in \cite{Deger:2019jtl} and could be a good starting point to understand this problem and generalize dynamical systems approach to a richer but relatively less challenging set up.

The second curious aspect is the intersecting RG flows that we observe on the phase picture for $1/2 < a^2 < 1$ (see Fig. \ref{fig:ag12}). Technically these come from zeroes of the scalar potential. Again from the point of view of a dynamical system this is perfectly fine and an example of a system with intersecting phase curves is the pendulum, where curves for counterclockwise and clockwise rotation intersect. It is not clear what is the meaning of such intersecting points from the QFT point of view. One possible explanation is that, these are simply an artifact of the change of variables. Unfortunately, we were not able to find such a dynamical system representation of the supergravity equations of motion, where singular points do not appear. Another possibility is that such points indeed represent some non-perturbative effect in the dual QFT that develops ersatz critical points. It is worth to mention that according to the phase picture at Fig. \ref{fig:ag12} there are multiple RG flows between the UV critical point $p_1$ and the IR stable point $p_4$, only one of which is supersymmetric. This is precisely the solution found in \cite{Deger:2002hv} that corresponds to the true superpotential and that is the $W_+$-type solution in this case. Although, the flows intersect, supersymmetry does not restore when going along a non-supersymmetric RG flow following a $W_-$ solution (say, green on Fig. \ref{fig:ag12}). This is simply because the superpotentials corresponding to the supersymmetric (orange) and non-supersymmetric (green) flows are different everywhere except at the critical points as can be seen from Fig. \ref{fig:WW0g12}.

The $D=3, \, \mc{N}=(2,0)$ gauged supergravity that we considered in this paper also allows compact target spaces for the sigma model \cite{Deger:1999st} and it will be interesting to apply our investigation to such a case. The supersymmetric RG flows for the $SU(2)/U(1) =\SS^2$ scalar coset space were studied in \cite{Deger:2002hv} where the true superpotential is given by a periodic function of the scalar field, namely $W_{\rm susy}=\cos \phi$ and here one might expect limit RG cycles. Moreover, the scalar potential has Minkowski and de Sitter vacua in addition to AdS and it would be interesting to look at flows between these critical points. Such an analysis could also provide valuable information for the dual 2-dimensional $\mc{N}=(2,0)$ CFTs \cite{Gukov:2015qea}.

Another interesting direction is to look for a special class of vacuum solutions given by gravitational instantons as in \cite{deHaro:2006ymc}, that satisfy vacuum Einstein equations and whose energy-momentum tensor vanishes. We checked that there are no such solutions in the theory considered here and hence one should consider generalizations of this model, such as by including Fayet-Iliopoulos terms as in \cite{Abou-Zeid:2001inc}. This is reserved for future work.

Given the results of \cite{Fortin:2012hn,Luty:2012ww} where the constraints coming from the a-theorem for a unitary $D=4$ QFT are inconsistent with RG cycles, it should be interesting to apply the dynamical system approach to five-dimensional gauged supergravity models. A suitable such theory is given in 
 \cite{Bobev:2020lsk} where 
 the full $D=5$, $\mc{N}=8$ gauged supergravity interacting with 42 scalars parametrizing the coset $\rmE_{6(6)}/\rmSU(8)$ is truncated to a model with only 2 scalar fields. 
 A supersymmetric domain wall solution to the truncated field equations describes the RG flow between $\mc{N} = 4$, $D=4$ SYM in the UV to the Leigh-Strassler deformed $\mc{N}=1$ theory in the IR, where both critical points are AdS \cite{Bobev:2020lsk}. It is clearly desirable to study the full RG flow phase picture for this theory. Another relevant model has been constructed in \cite{Karndumri:2022dtb}, that is, the $D=5$, $\mc{N}=4$ gauged supergravity with $\rmSO(2)\times \rmSO(3) \times \rmSO(3)$ gauge group, where the 25-dimensional scalar manifold is truncated to a 4-dimensional submanifold. We hope to come back to these issues in the near future.

\section*{Acknowledgments}

The authors would like to thank Ioannis Papadimitrou for helpful discussions, Irina Aref'eva for comments on the text and the anonymous referee for valuable comments and corrections. EtM thanks Bo\v{g}azi\c{c}i University for warm hospitality, where part of this work was carried. This work was supported by the Foundation for the Advancement of Theoretical Physics and Mathematics “BASIS”, grant No 21-1-2-3-1, by Russian Ministry of education and Science and by Tubitak Bideb-2221 fellowship program. The work of AG  presented in Section 3.2 was supported by Russian Science Foundation grant RSCF-22-72-1012.

\appendix
\setcounter{equation}{0}
\setcounter{equation}{0} \renewcommand{\theequation}{A.\arabic{equation}}

\section{Boundary value problem for scalars on AdS}
\label{appendixA}
In this appendix we briefly review the boundary value problem for scalar fields on AdS space-time. We first analyze a massive scalar field on a fixed AdS background in \ref{app:scalarads} and \ref{app:renorm} and then consider coupling with gravity in \ref{app:covren}.

\subsection{Boundary actions}
\label{app:scalarads}

As the simplest model of the holographic correspondence consider a massive scalar field in the 
AdS$_{\rm d+1}$ space-time whose action is
\begin{equation}
    S_0 = \fr12\int d^dx dz \sqrt{-g} (g^{\m\n} \dt_\m \f \dt_\n \f + M^2 \f^2),
\end{equation}
where the metric is defined by
\begin{equation}
    ds^2 = \fr{1}{z^2}(dz^2 + d\vec{x}^2).
\end{equation}
With properly chosen boundary conditions to be discussed below equations for the scalar field read
\begin{equation}
    z^2 \dt_z^2 \phi - (d-1)z\dt_z \f - (M^2 - z^2 \Box)\f = 0,
\end{equation}
where $\Box = \vec{\nabla}^2$. A general solution can be written in the following form
\begin{equation}
    \label{eq:solbessel}
    \f(z,\vec x ) = \int \fr{d^d k}{(2\p)^d} e^{- i \vec k \vec x} z^{\fr d 2} a(\vec{k})K_\n( k z),
\end{equation}
where $K_\n(kz)$ is the modified Bessel function and 
\begin{equation}
    \n = \sqrt{M^2 + \fr{d^2}{4}}.
\end{equation}
We see that well-behaved (stable) solutions exist only for the masses that satisfy Breitenlohner-Freedman bound
\begin{equation}
    M^2 \geq - \fr{d^2}{4}.
\end{equation}
Leading terms in the expansion of the Bessel function near $z=0$ behave as $z^\n$ and $z^{-\n}$ giving the standard representation of the near boundary behavior of the scalar field
\begin{equation}
    \label{eq:phi0A}
    \phi(z,\vec x ) = z^{\D_-} \f_0(\vec x ) + z^{\D_+} A(\vec{x}),
\end{equation}
where $\D_\pm$ are solutions to 
\begin{equation}
    \D(\D-d) = M^2.
\end{equation}
The standard prescription is to set $\phi_0$ as the source for an operator $\mc{O}(\vec x)$ of the boundary CFT, whose conformal weight is $\D_+$. Then the expectation value is 
\begin{equation}
    A(\vec x) = \langle \mc{O}(\vec x) \rangle_{\f_0},
\end{equation}
where the subscript means that the quantum average is taken in a theory with a source. To arrive at this prescription one notes that the first term in \eqref{eq:phi0A} dominates when $z \to 0$ leading to a finite on-shell action. Then the holographic correspondence gives
\begin{equation}
    \label{eq:adscft}
    e^{-S_{on-shell}[\phi_0]} = \langle e^{\int d^d x \mc{O}(\vec x) \phi_0(\vec x) } \rangle.
\end{equation}

The story is however more complicated and the picture above is actually a small piece of it corresponding to only Dirichlet boundary conditions, i.e., to the conditions when one fixes behavior of the scalar field on the boundary and sets $\d \f(z=0,\vec{x}) = 0$ when deriving equations of motion. Other boundary conditions, such as Neumann and mixed, allow to swap the meaning of $\phi_0$ and $A(\vec x)$, assigning to the latter the meaning of the source of an operator $\tilde{\mc{O}}$ of weight $\D_-$ in an alternative boundary CFT. To see the possibilities let us investigate the space of allowed conformal weights depending on the mass $M^2$
\begin{figure}[http]
    
    \centering
    \includegraphics[height=5cm]{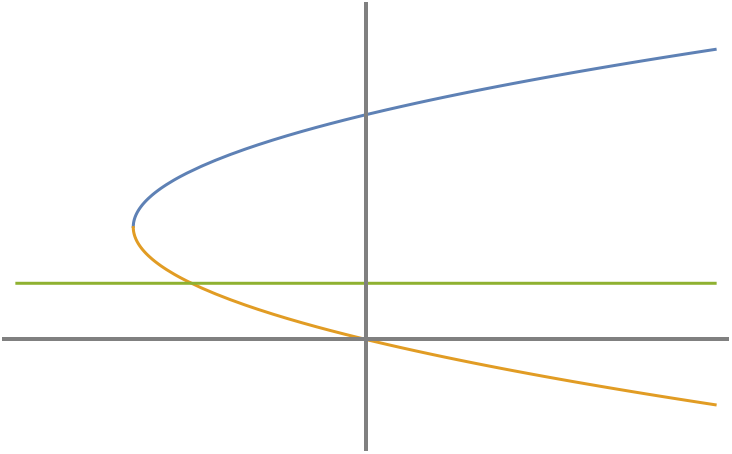}
    \caption{The space of possible conformal weights. The horizontal line depicts the unitarity bound $\D \geq \fr d2 -1$.}
    \begin{tikzpicture}[overlay, scale=0.625]
        \draw (3,9.2) node (p) {$\D_+$};
        \draw (3,3.2) node (m) {$\D_-$};
        \draw (5,6) node (u) {$\dfr d2-1$};
        \draw (6.5,4) node (m2) {$m^2$};
        \draw (-4.3,3.5) node (p) {$-\fr{d^2}{4}$};
        \draw (-2.8,3.5) node (p) {$-\fr{d^2}{4}+1$};

        \draw[dashed,thick] (-4.1,4.1) -- (-4.1,8);

         \draw[dashed,thick,red] (-3,4.1) -- (-3,8);

    \end{tikzpicture}
    
\label{fig:delta}
\end{figure}
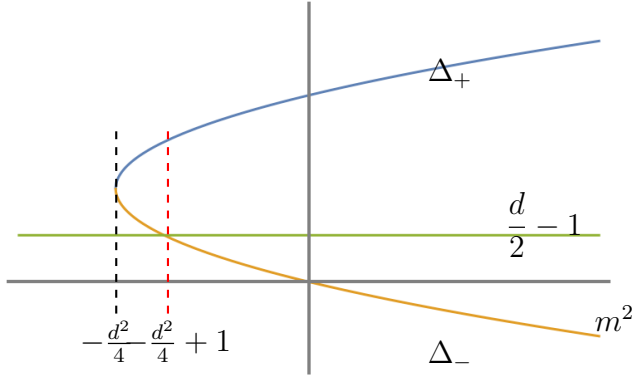

From Fig. \ref{fig:delta} one finds that in order to stay above the unitarity bound one is free to choose both $\D_+$ or $\D_-$ as the conformal weight for the dual operators, when 
\begin{equation}
    M^2 < -\fr{d^2}{4}+1.
\end{equation}
This is the area to the left of the right dashed line. To the right of the line only one branch of conformal weight is available and hence a scalar field of mass $M^2 > -\fr{d^2}{4}+1$ can only correspond to an operator of conformal weight $\D_+$. This has first been observed in \cite{Klebanov:1999tb}.

The actual behavior of the scalar field near the boundary and hence dimension of the corresponding operator depends on the choice of boundary conditions. In details this has been investigated in \cite{Minces:1999eg}, let us briefly recall their results. The main idea is that variation of the action on the boundary must be treated carefully and a proper boundary term must be added to end up with the standard equations of motion
\begin{equation}
    S= S_0 + S_B.
\end{equation}
Hence, for the variation we have
\begin{equation}
    \begin{aligned}
        \d S_0  & = S_0[\f + \d \f] - S_0[\f]  \\
        & =\int d^d x d z \sqrt{-g}\, \d \f \big(  - \Box_g + M^2\big)\f \\
        &+ \int d^d x \int_\e^{+\infty} dz \dt_z \Big(z^{-d+1} \dt_z \f \, \d \f\Big) \\
    \end{aligned} 
\end{equation}
where  $\e \ll 1$ has been introduced to regularize possible divergences near the boundary. In the usual case of Dirichlet boundary conditions one fixes value of the field $\f(0,\vec{x})$ one the boundary and hence imposes $\d \f (0,\vec{x}) = 0$. In general however this is not a must and e.g. for Neumann boundary conditions one fixes $\d \dt_z \f (0,\vec{x}) = 0$ instead. Note, that this is additional boundary data and hence $ \d \dot \f \neq \dt_z \d \f$ on the boundary. To properly take into account other boundary conditions than Dirichlet one has to add boundary terms to the full action, that implies for its variation 
\begin{equation}
    \d S =  - \int d^d x \e^{-d+1} \dt_z \f \, \d \f +\d S_B + \int \mathrm{EoM}.
\end{equation}
Depending on boundary conditions either $\d \f$, $\d \dt_z \f$ or their combination is fixed on the boundary. We have the following options:

\textbf{Dirichlet}: $\f(\e,\vec{x})$ is fixed, hence $\d \f (\e) = 0$ meaning $S_B = 0$. The on-shell action reads
\begin{equation}
    \begin{aligned}
        S_D  &= \fr12 \int d^d x dz \dt_\m (\sqrt{-g}g^{\m\n}\dt_\n \f\, \f)\\
        &=-\frac{1}{2}\int d^d x \e^{-d+1} \dt_z \f(\epsilon) \, \f(\epsilon)
    \end{aligned}
\end{equation}

\textbf{Neumann}: $z\dt_z\f(\e,\vec{x})$ is fixed, hence $\d \dt_z \f (\e) = 0$ meaning 
\begin{equation}
    \begin{aligned}
        S_B  & = -\int d^d x dz \dt_\m (\sqrt{-g}g^{\m\n}\dt_\n \f\, \f)\\
        &=\int d^d x \e^{-d+1} \dt_z \f(\epsilon) \, \f(\epsilon).        
    \end{aligned}
\end{equation}
The on-shell action reads
\begin{equation}
    \begin{aligned}
        S_N &= -\fr12 \int d^d x dz \dt_\m (\sqrt{-g}g^{\m\n}\dt_\n \f\, \f)\\
        &=\frac{1}{2}\int d^d x \e^{-d+1} \dt_z \f(\epsilon) \, \f(\epsilon),      
    \end{aligned}
\end{equation}
that is the opposite of the Dirichlet on-shell action.

\textbf{Mixed}: $\f(\e,\vec{x})+ \a z\dt_\z\f(\e,\vec{x})$ is fixed, hence $\d \f(\e) + \a \d \dt_z \f (\e) = 0$. The surface term should be 
\begin{equation}
    S_{B}=-\frac{\alpha}{2}\int d^d x \e^{-d+2}\dt_z \f(\epsilon)\dt_z \f(\epsilon).    
\end{equation}
The on-shell action reads
\begin{equation}
    \begin{aligned}
        S_M & = \fr12 \int d^d x dz \dt_\m \Big(\sqrt{-g}g^{\m\n}\dt_\n \f\, \big(\f + \a z \dt_z \f\big)\Big)\\
        &=-\frac{1}{2}\int d^d x \e^{-d+1}\big(\f(\epsilon)+\alpha\epsilon\dt_{z}\f(\epsilon)\big)  \,\dt_{z} \f(\epsilon),
    \end{aligned}
\end{equation}
that is the opposite of the Dirichlet on-shell action.

\subsection{Renormalization}
\label{app:renorm}

The boundary actions above are in general divergent in the limit $\e \to 0$, that requires a proper regularization procedure with further correct subtraction of divergent terms. In the case of the scalar field on a the fixed AdS background one can simply follow the standard procedure, that is to keep the leading divergent term and to rescale the boundary value by a proper power of $\e$. However, in a more general case of asymptotically locally AdS spaces this approach does not work. The reason is that the counterterms to the regularized action are defined on a surface $\S_\e$, that is close to the boundary and is diffeomorphic to it. In general the conformal symmetry present on the boundary breaks on the regularizing surface $\S_\e$ and hence not any boundary condition that makes sense on $\S_\e$ would do so on the conformal boundary. Apparently, one must define boundary data in terms of tensors w.r.t. the conformal group of the boundary. The idea is to decompose near-boundary bulk fields under the action of the dilatation operator and to keep a certain term in the expansion canceling the others by a proper counterterms action. It is most suggestive to observe the procedure on the example of the scalar field on the fixed AdS background. 

First, let us show that the desired term is the one that cannot be written in terms of a local functional of the boundary data. Let us look at solutions of the form
\begin{equation}
    \F(z, \vec{x}) = z^{\D_-}\f(z,\vec{x}).
\end{equation}
Hence, we have the following equation
\begin{equation}
     z^2 \dt_z^2 \phi +(d-2\D +1)z\dt_z \f - z^2 \Box \f = 0,
\end{equation}
where we denote $\D=\D_+$. Let us now try to solve this equation iteratively in the limit $z\to 0$, that means we expand $\f(z,\vec x)$ in powers of $z$ 
\begin{equation}
    \f(z,\vec x) = \sum_{n\geq 0} z^{2n}\f_{2n}(\vec x) ,
\end{equation}
and solve the resulting polynomial equation
\begin{equation}
    \sum_{n\geq 0}z^{2n+2} \Big( 2(n+1)(d -2 \D + 2n+2) \f_{2n+2}- \Box \f_{2n}\Big)=0.
\end{equation}
We see that for all terms with $n < \n=\D-\frac{d}{2}$ we may express all $\f_{2n}(\vec x)$ as local functionals of $\f_0$:
\begin{equation}
    \label{eq:lowernu}
    \f_{2n}(\vec x) = \fr{1}{4n(n-\n)}\Box \f_{2(n-1)}(\vec x).
\end{equation}
Assuming $\n \in \ZZ_+$ the above procedure breaks at $n=\n$ and hence $\f_{2\n}(\vec x)$ cannot be written in the above form, instead it is a non-local functional of the boundary data as we see below. To see that this is the only data one needs let us deform the expansion as follows
\begin{equation}
    \f(z,\vec x) = \sum_{n\geq 0} z^{2n}\f_{2n}(\vec x) + z^{2\n} \log z \y_{2\n}(\vec x) .
\end{equation}
In this case for $\f(2n)$ with $n<\n$ we have \eqref{eq:lowernu}, the function $\y(\vec x)$ is then determined as follows
\begin{equation}
    \y_{2\n}(\vec x) = \fr{1}{(2\n-1)^2}\Box \f_{2(\n-1)}(\vec x).
\end{equation}
All further functions $\f_{2n}$ with $n > \n$ are written again as local functionals of $\y(\vec x)$ and $\f_{2\n}$, e.g. we have
\begin{equation}
    \f_{2(\n+1)}(\vec x) = \fr{1}{4(\n+1)} \Big(\log z \Box \y(\vec x) + \Box \f_{2\n}(\vec x)\Big).
\end{equation}
Hence, only $\f_{2\n}(\vec x)$ remains undetermined as a local functional by the bulk equations. This is precisely the boundary data one has to define. Indeed, to renormalize the boundary action one must add counterterms $S_{ct}$ that are local functionals of the bulk fields. Apparently, $S_{ct}$ may contain only fields $\f_{2n}(\vec x)$ with $n < \n$, leaving $\f_{2\n}(\vec x)$ in the boundary action. Taking the limit $z \to 0$ one finds out that such a renormalized action
\begin{itemize}
    \item is finite;
    \item is conformally invariant;
    \item reproduces the standard results in the AdS case.
\end{itemize}
    
The above arguments apparently do not apply when $\n \notin \ZZ_+ $, in which case one should use slightly more subtle techniques. Let us now proceed with such more general arguments that will end up with the definition of the so-called renormalized radial momentum $\hat\p_+$. For that lest us start with a general solution \eqref{eq:solbessel}
\begin{equation}
    \label{eq:solbessel1}
    \f(z,\vec x ) = \int \fr{d^d k}{(2\p)^d} e^{- i \vec k \vec x} z^{\fr d 2} a(\vec{k})K_\n( k z),
\end{equation}
where $K_\n( k z)$ is the modified Bessel function and as before $\n^2 = M^2 + d^2/4$. We are interested in the boundary action that is written in terms of the scalar field or its momentum $\p_\f = \dt_z \f$ and the boundary data, that is the value of the scalar field on the boundary (Dirichlet boundary conditions), the value of the momentum (Neumann boundary conditions) or some combination of these (mixed boundary conditions). Hence, in general we are interested in defining the (renormalized) momentum $\p_\f$ in terms of the boundary data $\f(\e,\vec x )$, where $\e \to 0$ is some regulating parameter. From this we determine Fourier coefficients $a(\vec{k})$:
\begin{equation}
    a(\vec k) = \e^{-\fr d2} \fr{1}{K_\n(k\e)} \int d^d x \,\f(\e,\vec x) \, e^{i \vec k \vec x}.
\end{equation}
For the momentum $\p = \sqrt{g}g^{zz}\dt_z \f$ this gives
\begin{equation}
    \label{eq:pi}
    \p(\e,\vec x) = \e^{-d} \int  \fr{d^d yd^d k}{(2\p)^d}e^{-\vec{k}(\vec x - \vec y)} \left[\fr d2 + k\e \fr{K'_\n(k\e)}{K_\n(k\e)}\right]\f(\e,\vec y).
\end{equation}
Eliminating the momentum integral we arrive at logarithmic derivative of $K_\n(\e \sqrt{\Box})$, that has expansion of the same form as in \eqref{eq:lowernu} as we show below. Hence it is suggestive to proceed with a series representation of the modified Bessel functions for small argument:
\begin{equation}
    \begin{aligned}
        & \n \notin \ZZ_+ :\\
        &  K_\n(\z) = \fr{\p}{2 \sin(\p \n)}\SSum_{n\geq 0}\fr{1}{n!}\left[\fr{1}{\G(n-\n + 1)}\left(\fr{\z}{2}\right)^{2n-\n}\right.\\
        &\left. \qquad- \fr{1}{\G(n+\n + 1)}\left(\fr{\z}{2}\right)^{2n+\n}\right];\\
        & \n \in \ZZ_+ : \\
        & K_\n (\z) = \SSum_{n=0}^{\n-1}(-1)^\n \fr{\G(\m-n)}{n!}\left(\fr{\z}{2}\right)^{2n-\n}\\
        &    -  \SSum_{n\geq0}\left[\log\fr{\z}{2} - \fr{\lambda(\n,n)}{2}\right]\fr{(-)^\n}{n!}\G(n+\n+1)\left(\fr{\z}{2}\right)^{2n+\n},
    \end{aligned}
\end{equation}
where 
\begin{equation}
    \lambda(\n,n) = -2\g + \Sum_{m=1}^{n}\fr1m + \Sum_{m=1}^{n+\n}\fr1m ,
\end{equation}
and for $K_0(\z)$ one drops the first sum in the second line and sets $\n=0$ in the remaining terms. Substituting the expansion for say $\n \notin \ZZ_+$ into \eqref{eq:pi} we obtain
\begin{equation}
    \begin{aligned}
        \p(\e,\vec{x}) & = \sqrt{\g}\big(\p_{(\D_-)} + \p_{(\D_-+2)} + \dots + \p_{(\D_+)} \\
        &+ \tilde{\p}_{(\D_+)}\log z^2 + \dots \big),\\
        \p_{(\D_-)} & = \D_- \int_{ky} e^{-i\vec{k}(\vec{x}-\vec{y})}\f(\e,\vec{y}), \\
        \p_{(\D_-+2)} & = -\fr{1}{2(\n-1)}\e^2 \int_{ky} e^{-i\vec{k}(\vec{x}-\vec{y})}k^2\f(\e,\vec{y}),\\
        \p_{(\D_-+4)} & = \fr{\n^2 - 2\n +2}{16(\n-2)(\n-1)^2} \e^4\int_{ky} e^{-i\vec{k}(\vec{x}-\vec{y})}k^4\f(\e,\vec{y}),\\
        &\vdots \\
        \p_{(\D_+)} & = -\fr{2 \Gamma(1-\n)}{2^{2\n}\Gamma(\n)} \e^{2\n}\int_{ky} e^{-i\vec{k}(\vec{x}-\vec{y})}k^{2\n}\f(\e,\vec{y}),
    \end{aligned}
\end{equation}
where $\g = \e^{-d}$ is determinant of the metric induced on the surface $\S_\e$ and we denote
\begin{equation}
    \int_{ky} = \int \fr{d^d y d^d k}{(2\p)^d}. 
\end{equation}
We see that all $\p_{\D_- +2n}$ with $n < \n$ are local functionals of the value $\f(\e,\vec x)$ of the bulk scalar field on the boundary and have the form
\begin{equation}
    \p_{(\D_- + 2n)} \sim \Box^{n} \f(\e,\vec x), \quad n < \n.
\end{equation}
First term that is a non-local functional of $\f(\e,\vec x)$ is $\p_{(\D_+)} \sim \Box^\n \f(\e,\vec x)$. This is precisely what we have observed before for integer $\n$ when solving Klein-Gordon equation iteratively. Strictly speaking, the above is true only for non-integer $\n$, however the same derivations can be repeated for $\n \in \ZZ_+$, where $\p_{(\D_+)}$ will be the first term containing $\log k$, and hence again non-local.

The next step is to renormalize the action, that is to add $S_{ct}$, a local functional of bulk fields, such that to  keep only $\p_{(\D_+)}$. Subtracting all $\p_{(\D_- + 2n)}$ will be a correctly defined procedure as these are eigenfunctions of the dilation operator and have well defined behavior as $\e \to 0$. Hence, we write
\begin{equation}
    \d S_{ren} = \d S + \d S_{ct} = \int_{\S_\e} \p_\f \d\f + \d S_{ct}.
\end{equation}
Defining $\d S_{ct} = \sum_{n < \n} \p_{(\D_- + 2n)} \d \f$ we end up with the following variation of the renormalized action
\begin{equation}
    \d S_{ren} = \int_{\S_\e} \sqrt{\g}\p_{(\D_+)} \d \f(\e,\vec x) + \dots,
\end{equation}
where dots denote terms of higher order in $\e$. Now, since the leading behavior of the scalar field near the boundary is 
\begin{equation}
    \d \f (\e,\vec x) = \e^{\D_-} \d \f_-(\vec x),
\end{equation}
we have
\begin{equation}
    \d S_{ren} = \int_{\S_\e} \e^{-d + \D_-}\p_{(\D_+)} \d \f_-(\vec x) + \dots.
\end{equation}
Finally, defining $\hat \p_+ = \e^{-\D_+}\p_{(\D_+)}$ at taking the limit $\e \to 0$ we obtain
\begin{equation}
    \d S_{ren} = \int_{\S_\e} \hat\p_{+} \d \f_-(\vec x).
\end{equation}
Explicitly for the renormalized radial momentum we have
\begin{equation}
    \begin{aligned}
        &\hat\p_+  = \e^{-\D_+}\p_{(\D_+)} \\
        &= \e^{-\D_+ + \D_- + 2\n}\fr{2 \G(1-\n)}{2^{2\n}\Gamma(\n)}\int \fr{d^d y d^d k}{(2\p)^d} e^{-i\vec k (\vec x - \vec y)}k^{2\n}\f_-(\vec y)\\
        & = \fr{\Gamma(1-\n)}{2^{2\n-1}\Gamma(\n)}(-\Box)^\n \f_-(\vec x).
    \end{aligned}
\end{equation}

Renormalized on-shell action for various boundary conditions then reads. 
\begin{equation}
    \begin{aligned}
         S_D[\f_-] &  = \fr{\G(1-\n)}{4^\n l \G(\n)} \int_x \f_-(x)(-\Box)^\n \f_-(x) \\
         &\sim \int_{xy} \fr{\f_-(x) \f_-(y)}{|x-y|^{2 \D_+}} ,\\
         S_N[\hat\p_+] & = -\fr{4^\n l \G(\n)}{4 \G(1-\n)} \int_x \hat\p_+(x)(-\Box)^{-\n} \hat\p_+(x)\\
         &\sim \int_{xy} \fr{\hat\p_+(x) \hat\p_+(y)}{|x-y|^{2 \D_-}} ,\\
         S_M[J_\x] & = -\fr{1}4 \int_x J_\x(x)\left[\x -\fr{\G(1-\n)}{4^\n l \G(\n)} (-\Box)^\n\right]^{-1} J_\x(x) \\
         & \sim \int_{xy} {J_\x(x) G_\x(x-y)J_\x(y)} ,\\
    \end{aligned}
\end{equation}
where $G_\x(x-y)$ is a Green's function of the operator $\Box^\n - \x$. We see that in this case Dirichlet boundary conditions correspond to the $\D_+$ branch of conformal dimensions, Neumann -- to the $\D_-$ branch. Mixed boundary conditions correspond to adding multi-trace deformations.

We see that to go from the Dirichlet boundary conditions, when one fixes $\f_-(\vec x)$ to Neumann boundary conditions when $\hat \p_+$ is fixed instead one formally writes
\begin{equation}
    \f_-(\vec x) = \fr{2^{2\n -1}\G(\n)}{\G(1-\n)}(-\Box)^{-\n} \hat \p_+(\vec x ).
\end{equation}
The action $S_D[\f_-]$ then becomes $S_N[\hat \p_+]$. The general near boundary behavior of the scalar field in the AdS space-time is given by
\begin{equation}
    \f(z,\vec x) = z^{\D_-}\Big(\f_-(\vec x) + \mc{O}(z^2)\Big) + z^{\D_+}\Big(A(\vec x) + \mc{O}(z^2)\Big).
\end{equation}
Comparing with the effective actions above we see that in the case of Dirichlet boundary conditions $\f_-(\vec x)$ gives the source $J(\vec x)$ of an operator of conformal dimension $\D_+$, while in the Neumann case the source is given by $A(\vec x) = \hat \p_+$. Renormalized one-point correlation functions in the presence of sources then read
\begin{equation}
\label{eq:corrapp}
    \begin{aligned}
        & D: & \langle \mc{O}_{\D_+} \rangle^{ren}_s &= \fr{\d S_D[\f_-]}{\d \f_-} = \fr{\G(1-\n)}{2^{2\n - 1}\G(\n)} (-\Box)^\n \f_-  \\
        &&&= \hat \p_+, \\ 
        & N: & \langle \mc{O}_{\D_-} \rangle^{ren}_s &= \fr{\d S_N[\hat\p_+]}{\d \hat \p_+} = \fr{2^{2\n-1}\G(\n)}{\G(1-\n)} (-\Box)^{-\n} \hat \p_+ \\
        &&&= \f_-, \\ 
    \end{aligned}
\end{equation}

\subsection{Scalars coupled to gravity}
\label{app:covren}

Backgrounds given by an asymptotically locally anti de Sitter (AlAdS) space-time  are solutions to the gravity theory with a supporting scalar field. For the minimal coupling the action can be written as follows
\begin{equation}
    S = \int_{\mc{M}}d^{d+1} x \sqrt{-g}\bigg[\fr{-1}{2 \k^2} R + \fr12 g^{\m\n}\dt_\m \f \dt_\n \f + V(\f)\bigg] + S_B,
\end{equation}
where $S_B$ is a boundary action, whose particular form depends on boundary conditions on $\S_w = \dt \mc{M}$. Let us first consider variation of the usual Einstein-Hilbert term
\begin{equation}
    \begin{aligned}
        \d S_{EH} =& - \fr1{2\k^2} \int d^{d+1}x \sqrt{g}  \Big[R_{\m\n} - \fr12 g_{\m\n}R\Big] \d g^{\m\n} \\
        &- \fr{1}{2\k^2} \int d^{d+1}x \sqrt{g}\, g^{\m\n}\d R_{\m\n}.
    \end{aligned}
\end{equation}
The first term gives the Einstein tensor in the LHS of gravity equations, while the last term can be written as a total derivative:
\begin{equation}
    g^{\m\n}\d R_{\m\n} = g^{\m\n}\Big[\nabla_\m \d \G_{\n \r}{}^\r - \nabla_\r \d \G_{\m\n}{}^\r\Big].
\end{equation}
The crucial observation here is that the tensor $\d \G_{\m\n}{}^\r$ involves not only variations $\d g_{\m\n}$, but also variations of its derivative $\d \dt_\m g_{\r\s}$. In the most general case when the space has a boundary variation and derivative can not be interchanged, i.e. $\d \dt_\m g_{\r\s} = 0$ on a boundary does not follow from $\d g_{\m\n} = 0$ on the same boundary. These are different boundary conditions and hence one must add a boundary action to cancel the contribution from $\d R_{\m\n}$. This is the so-called Gibbons-Hawking-York term
\begin{equation}
    S_B = -\fr{1}{\k^2} \int_{\S_w} d^d x \sqrt{\g} K,
\end{equation}
where $K = \g^{ij}K_{ij}$ is the trace of the extrinsic curvature 
\begin{equation}
    K_{ij} = \fr12 \dot{\g}_{ij} ,
\end{equation}
of the boundary $\S_w$. 

Similarly for the variation w.r.t. the scalar field one has on-shell
\begin{equation}
    \label{eq:dS}
    \d_\f S = \int_{\S_w} \p_\f \d \f, 
\end{equation}
where $\p_\f = \sqrt{g}\dot \f$ denotes the canonical momentum. Now as in the previous subsection for Dirichlet boundary conditions one would simply set $\d \f =0$, while in more general case additional care must be taken. However, even before that one notes that the actual boundary data lives on the conformal boundary of an AlAdS space-time, while $\S_w$ is in contrast an ordinary boundary, that breaks conformal invariance for any finite $w$. Hence not any condition that makes sense on $\S_w$ will do so on the conformal boundary $\S_\infty$. To ensure that one must formulate boundary data on $\S_w$ in terms of quantities, that transform covariantly under conformal transformations. For that it has been suggested in \cite{Papadimitriou:2004ap} to use the observation that the derivative $\dt_w$ for only the leading asymptotic behavior as $w\to \infty$ can be replaced by dilatation operator
\begin{equation}
    \dt_w = \int_{\S_w}d^dx\,\dt_w \f \,\fr{\d}{\d \f} \sim \d_D,
\end{equation}
where 
\begin{equation}
    \d_D = - \D_-  \int_{\S_w} \f \fr{\d}{\d \f}.
\end{equation}
Hence, trading $\dt_w$ for the dilatation operator and keeping only terms of particular weight under $\d_D$ in \eqref{eq:dS} we use only quantities that are well defined in the limit $w \to \infty$. For that decompose the fields in a sum of eigenfunctions of the dilatation operator \cite{Papadimitriou:2004rz}. For the scalar field this yields the standard expression
\begin{equation}
    \f(w,\vec{x}) \underset{w \to \infty}{\sim} e^{-\D_- w} \big(\f_-(\vec{x})  + \dots ) + e^{-\D_+ w}\big(\f_+(\vec{x}) + \dots\big).
\end{equation}
Given the definition $\p_\f = \sqrt{\g}\dot{\f}$ for the canonical momentum one has
\begin{equation}
        \p_\f = \sqrt{\g} \Big(\sum_{\D_- \leq s < \D_+} \p_{(s)} + \p_{(\D_+)} + \tilde{\p}_{(\D_+)}\log e^{-2 w} + \dots\Big),
\end{equation}
where 
\begin{equation}
    \begin{aligned}
        \d_D \p_{(s)} & = - s \p_{(s)}, && \D_- \leq s < \D_+, \\
        \d_D \p_{(\D_+)} & = - \D_+ \p_{(s)} - 2 \tilde{\p}_{(\D_+)}, && , \\
        \d_D \tilde{\p}_{(\D_+)} & = - \D_+ \tilde{\p}_{(\D_+)}.
    \end{aligned}
\end{equation}
Hence, we see that $\p_{(s)}$ for $\D_- \leq s < \D_+$ can be determined as local expressions of the corresponding components of $\phi(0,\vec{x})$, while $\p_{(\D_+)}$ cannot as it transforms inhomogeneously. Precisely this component corresponds to VEV of the dual operator in the case of Dirichlet boundary conditions. This implies that $\p_{(\D_+)}$ must be fixed by boundary data, while the other contributions can be canceled out by proper boundary counterterms, whose variation read
\begin{equation}
    \d S_{ct} = - \sum_{s < \D_+}\int d^d x \sqrt{\g} \, \p_{(s)} \d \f. 
\end{equation}
Hence, finally we have for the variation of the full matter on-shell action is
\begin{equation}
    \d S + \d S_{ct} = \int_{\S_w} d^dx \sqrt{\g} \p_{(\D_+)} \d \f.    
\end{equation}
The integral is a scalar under $\d_D$ and hence behaves well in the limit $w \to \infty$, implying that boundary conditions must be formulated in terms of $\p_{(\D_+)}$ to be well defined in this limit. To see that define
\begin{equation}
    \hat{\p}_{+} = \lim_{w \to \infty} e^{\D_+ w} \p_{(\D_+)} ,
\end{equation}
and evaluate the variation on the conformal boundary
\begin{equation}
    \d (S+S_{ct}) = \int_{\dt \mc{M}} d^d x \sqrt{g_{(0)}} \hat{\p}_+ \d \f_-,
\end{equation}
where $\g = e^{d w} g_{(0)}$ and the limit is taken such that $\dt \mc{M} = \S_{\infty}$. We see that the boundary data is given by fixed values of $\hat{\p}_+$ and $\f_-$. Let us now give more details on the concrete cases of Dirichlet, Neumann and mixed boundary conditions.

For \textbf{Dirichlet} boundary conditions we have $\d \f_- = 0$, that fixes behavior of the scalar field at the boundary. In this case the renormalized variation of the action $\d(S+S_{ct}) = 0$ and no extra boundary term is needed. This is precisely as in the previous subsection.  

\textbf{Neumann} boundary conditions fix the value of $\hat\p_+$ on the boundary and hence the boundary action can be taken in the form
\begin{equation}
    S_B = -\int_{\dt \mc{M}} d^d x \sqrt{g_{(0)}} \f_- \hat{\p}_+.
\end{equation}
In this case the on-shell variation of the full action reads
\begin{equation}
    \d S_N = \d(S+S_{ct} + S_{B}) = -\int_{\dt \mc{M}} d^d x \sqrt{g_{(0)}} \f_- \d\hat{\p}_+ ,
\end{equation}
which vanishes when $\d \hat\p_+ = 0 $.

In the case of \textbf{mixed} boundary conditions one fixes the value of $\y = - \hat \p_+ - f'(\f_-)$ which implies the following boundary  term
\begin{equation}
    S_B = \int_{\dt \mc{M}} d^d x \sqrt{g_{(0)}} \big(- \hat\p_+ \f_- +f(\f_-) - \f_- f'(\f_-)\big) \, ,
\end{equation}
The on-shell variation of the full action is then
\begin{equation}
    \d S_M = \d(S+S_{ct} +S_B) =  \int_{\dt \mc{M}} d^d x \sqrt{g_{(0)}}  \f_- \d \y.
\end{equation}

\textbf{The crucial observation} here is that the correctly defined on-shell action is finite on the conformal boundary. Hence, the above definition coincides with older approaches based on the finiteness of the action. In the case of a scalar field on AdS space-time and mixed boundary conditions with quadratic $f(\f_-) = \x \f_-^2$ the above procedure can be performed exactly, as it has been shown in the previous subsection.

\section{Derivation of the autonomous system equations}
\label{app:autonom}
\setcounter{equation}{0}
\setcounter{equation}{0} \renewcommand{\theequation}{B.\arabic{equation}}
Here we give the derivation of the dynamical equations \eqref{eqZX} for $Z=(1+ e^{\f})^{-1}$ and $X = \dot \f/\dot A$. The starting point is the following expressions given in \eqref{eq:V2W} and \eqref{first} in the main text:
\begin{equation}
    \begin{aligned}
        V & = \fr{a^2}{4}W'^2 - \fr12 W^2, \\
        \dot A & = - W, \\
        \dot \f & =  a^2 W'.
    \end{aligned}
\end{equation}
The first one is basically the definition of the superpotential $W=W(\phi)$. Assuming that $A=A(r)$ is a monotonic function of the radial coordinate of the domain wall, that is the case for an RG flow solution, one is able to consider $A$ as the flow parameter. Hence, for the derivative of $Z=(1+ e^{\f})^{-1}$ we have
\begin{equation}
    \begin{aligned}
        \fr{dZ}{dA} = \fr{\dot Z}{\dot A} = - \fr{e^{\f}}{(1+e^\f)^2}\fr{\dot \f}{\dot A} = X Z(Z-1).
    \end{aligned}
\end{equation}
To proceed with the similar equation for $X$ we first list some useful identities
\begin{equation}
    \begin{aligned}
        \ddot{\f}& = a^2 W'' \dot \f = 2 a^2V' - 2 \dot{A} \dot \f,\\
        \ddot{A} & = - W' \dot \f = -\fr{\dot \f^2}{a^2}.
    \end{aligned}
\end{equation}
Now we write
\begin{equation}
    \label{eq:dXdA0}
    \begin{aligned}
        \fr{dX}{dA} = \fr{\dot X}{\dot A} = \fr{\ddot \f}{\dot A^2} - \fr{\dot \f \ddot A}{\dot A^3} = \fr{2a^2}{\dot A^2}V' - 2X + \fr{X^3}{a^2}.
    \end{aligned}
\end{equation}
To get rid of $\dot A^2$ in the denominator we again use the relation between the potential $V$ and the  superpotential to arrive at
\begin{equation}
    \label{eq:dotA2V}
    {4}V = \dot A^2 \left(\fr{X^2}{a^2} - 2\right).
\end{equation}
Hence, we finally get
\begin{equation}
    \label{eq:dXdA}
    \fr{dX}{dA} = \fr{a^2}{2}\fr{V'}{V}\left(\fr{X^2}{a^2} - 2\right) + X \left(\fr{X^2}{a^2} - 2\right),
\end{equation}
where the potential as well its derivative are functions of $Z$ only. Note, that the above is true for any scalar potential, given the domain wall ansatz.

\appendix

\newpage
\bibliography{bib.bib}
\bibliographystyle{utphys.bst}

\end{document}